\documentclass[a4paper,12pt]{article}
\usepackage[utf8]{inputenc}
\usepackage{amsfonts}
\usepackage{amssymb}
\usepackage{graphicx}
\usepackage{amsmath}
\usepackage[table,xcdraw]{xcolor}
\usepackage{tikz}
\usetikzlibrary{decorations.markings}
\usepackage{enumerate}
\usepackage{subfig}
\usepackage{color}
\usepackage{tikz}\usetikzlibrary{calc}
\usepackage{float}
\usepackage{here}
\usepackage{cite}
\usepackage{mathrsfs}
\usepackage{float,epsfig}
\usepackage{dcolumn}

\usepackage{graphicx}
\usepackage{bm}
\usepackage{amsmath,amssymb,amsthm}
\usepackage[colorlinks=true,linkcolor=blue,citecolor=red]{hyperref}
\newcommand{\be}{\begin{equation}}
\newcommand{\ee}{\end{equation}}
\newcommand{\bea}{\setlength\arraycolsep{2pt} \begin{eqnarray}}
\newcommand{\eea}{\end{eqnarray}}

\def\0{{\sst{(0)}}}
\def\1{{\sst{(1)}}}
\def\2{{\sst{(2)}}}
\def\3{{\sst{(3)}}}
\def\4{{\sst{(4)}}}
\def\5{{\sst{(5)}}}
\def\6{{\sst{(6)}}}
\def\7{{\sst{(7)}}}
\def\8{{\sst{(8)}}}
\def\sst#1{{\scriptscriptstyle #1}}

\definecolor{lime}{HTML}{A6CE39}
\newcommand{\orcidicon}{%
    \begin{tikzpicture}
    \draw[lime, fill=lime] (0,0)
        circle [radius=0.16]
        node[white] {{\fontfamily{qag}\selectfont \tiny ID}};
    \draw[white, fill=white] (-0.0625,0.095)
        circle [radius=0.007];
    \end{tikzpicture}   \hspace{-2mm}
}
\newcommand\orcidAdil{{\href{https://orcid.org/0000-0001-7623-5541}{\orcidicon}}}
\newcommand\orcidHasan{{\href{https://orcid.org/0000-0001-7408-0910}{\orcidicon}}}

\newcommand\orcidSedra{{\href{https://orcid.org/0000-0001-5160-7564}{\orcidicon}}}
\newcommand\orcidEssabani{{\href{https://orcid.org/0000-0001-5667-1071}{\orcidicon}}}

\newcommand\orcidSekhmani{{\href{https://orcid.org/0000-0002-4325-5252}{\orcidicon}}}

\makeatletter \@addtoreset{equation}{section}

\usepackage{multirow}
\begin{document}
%

\title{\normalsize
{\bf \Large	 Thermodynamic and Optical  Behaviors \\of Quintessential Hayward-AdS Black Holes }}
\author{ \small   A. Belhaj\orcidAdil\!\! $^{1}$\footnote{a-belhaj@um5r.ac.ma},  M. Benali\orcidAdil\!\!$^{1}$\footnote{mohamed\_benali4@um5.ac.ma}, H. El Moumni\orcidHasan\!\!$^{2}$\thanks{h.elmoumni@uiz.ac.ma},
	M. A. Essebani\orcidEssabani\!\!$^{3}$,  M. B. Sedra\orcidSedra\!\!$^{3,4}$,  Y. Sekhmani\orcidSekhmani\!\!$^{1}$\footnote{ Authors in alphabetical order.}
	\hspace*{-8pt} \\
	{\small $^1$ D\'{e}partement de Physique, Equipe des Sciences de la mati\`ere et du rayonnement, ESMaR}\\
{\small   Facult\'e des Sciences, Universit\'e Mohammed V de Rabat, Rabat,  Morocco} \\
	{\small $^{2}$  EPTHE, D\'{e}partement de Physique, Facult\'e des Sciences,   Universit\'e Ibn Zohr, Agadir, Morocco} \\
	{\small $^{3}$  D\'{e}partement de Physique,  Laboratoire de physique des Mat\'eriaux et Subatomique, LPMS}\\   {\small Facult\'{e}
		des Sciences, Universit\'{e} Ibn Tofail, K\'{e}nitra,
		Morocco } \\ {\small  $^4$ Moulay Ismail University, FSTE, LSTI, BP-509 Boutalamine,    Errachidia 52000,  Morocco    }
}

\maketitle

\vspace{-2.5em}
%
%

	\begin{abstract}
		{\noindent}
Motivated by   Dark Energy (DE) activities, we  study  certain physical behaviors of  the quintessential  Hayward-AdS black holes in four dimensions. We  generalize  some physical properties of  the ordinary     Hayward AdS black holes without  the dark sector. We elaborate a study in terms of the new quantities  $c$ and $\omega_q$ parametrizing the dark sector moduli space.   We    investigate  the effect of such  parameters on certain thermodynamic and optical  aspects.  To  show the  quintessential thermodynamic  behaviors, we first  reconsider the critical properties   of the ordinary solutions.   We  find that the equation of state predicts a universal  ratio  given by $\chi_0=\frac{P_cv_c}{T_c}=\frac{27-3\sqrt{6}}{50}$, which is  different than the universal  one  appearing  for Van der Waals fluids.   Considering  the quintessential solutions and  taking  certain values of the DE state parameter $\omega_q$,  we  observe  that the new ratio   depends on   the DE scalar  field intensity $c$.   In certain regions of the moduli space,  we show  that  this ratio can be factorized  using two terms   describing  the absence  and the presence of the dark sector.   Then, we  analyze  also   the DE effect on  the heat engines.   For the optical aspect, 
 we  study   the influence  of DE on the shadows  using one dimensional real curves. Finally, we discuss  the associated  energy emission rate, using the dark sector.
	\end{abstract}
\newpage

\tableofcontents

\section{Introduction}

Recently,  the  phase structure of  the  Anti de Sitter (AdS)  black holes has received more  attention  from  the extended phase space point of views  \cite{Kastor:2009wy,Dolan:2011xt}. In such an extended space,  it has been  implemented both the pressure and   the volume as thermodynamic  variables \cite{Kubiznak:2012wp}. Various interesting phenomena of  the AdS black holes have been explored, such as reentrant phase transitions\cite{Altamirano:2013ane,Kubiznak:2015bya}, triple points \cite{Altamirano:2013uqa},  and $\lambda$-line phase transitions \cite{Hennigar:2016xwd}. These activities  have    suggested   that the  AdS black holes offer huge similarities  with the  thermodynamic systems.\\
It has been remarked  that the investigation  of the  black hole singularities has always been a real crisis in general relativity theory. The singularities  are considered as serious problems  in such  a theory\cite{ma1,ma2,ma3,ma4,ma5,ma6,ma7}. To overcome such  issues,  many suggestions  have been  proposed.  In particular,     regular black hole solutions have been elaborated. Among others,   the theory of general relativity coupled to nonlinear electrodynamics  has been  also considered as  an interesting candidate \cite{Bronnikov:2000vy,Dymnikova:2004zc}. Alternative  ways  generate  regular solutions  containing a  critical scale, mass, and charge parameters restricted by certain  values, depending  only on the type of the curvature invariants \cite{Polchinski:1989ae}.   Hayward   presented a static spherically symmetric black hole being  near the origin behaves like a de Sitter space-time. Precisely, its curvature  is invariant  everywhere  and satisfies  the weak energy condition\cite{Hayward:2005gi}. Several  Hayward-like black holes have been  constructed  after  the original one by introducing   an  irregularity to topological changes.  This  offers a  possibility to build  spaces with a maximum curvature inside the black hole regions \cite{Frolov:1989pf,Mukhanov:1991zn,Frolov:2016pav}. Regular black hole interior solutions  have been also  found in Loop Quantum Gravity \cite{Modesto:2006mx}. In particular,  they represent  relevant ingredients  needed to understand  the associated  physical    theories.\\
In  addition to the  singularity problems, the general relativity theory should resolve certain questions  which remain without a   consensus.  These questions  may concern  the dark  sector associated with the  dark matter \cite{Bertone:2016nfn} and the dark energy of late-time cosmology\cite{Huterer:2017buf}.  As a way to investigate the  models dealing with such as a sector, a scalar field is usually introduced \cite{ Gannouji:2019mph,180}. Concretely, the quintessence remains the simplest and the promising one \cite{Caldwell:1997ii,Uniyal:2014paa,Konoplya:2019sns}. 
Recently,  it  has been shown that the  observational results could confirm  that our universe is expanding with
acceleration behaviors\cite{RF1,RF2}.  It has been remarked that  this surprisingly accelerated expansion  can be explained  by the  introduction of  the  dark energy, which accounts  about 70\% of the universe. It involves  a  negative pressure driving  the expansion of the universe.  Precisely, it   has been suggested that such an energy could be  modeled, in terms of  a  quintessential  scalar field   being considered as a spatially homogeneous real scalar field with intensity $c$. Treated as a perfect fluid with a pressure $p$ and  an energy density $\rho_q$, such a DE is controlled by the equation of state  $p=\omega_q \rho_q$  where   $\omega_q$ is called   a  state  parameter  with the constraint  $-1<\omega_q<-{1}/{3}$. \\Kiselev  first derived the solutions of the black hole in the presence of  the quintessence\cite{Kiselev:2002dx}.  Following this work, many black holes surrounded  by the quintessential  dark fields  have been  dealt with   by unveiling certain data of the associated physics\cite{Chen:2008ra,Chen:2012mva,Chabab:2017xdw,Belhaj:2020rdb}.\\
Beside  phase transitions and critical phenomena,  developments in  the black hole thermodynamics  have provided many works including  the Joule-Thomson expansion \cite{JT1,JT2}  and the  holographic heat engine  behaviors \cite{27,Belhaj:2015hha,28,NPBx}.   Considering  the AdS  black  holes as  heat engines,  various  black holes have  been examined.  Concretely,  the ordinary Hayward AdS black holes without external moduli space  have  been studied in \cite{Kubiznak:2012wp,29}.  The effect of DE on   engine  behaviors for RN-AdS black holes  have been  elaborated in  \cite{30}. It has been shown that the  quintessence,  controlled by   an  external  moduli  space,   could improve the associated efficiency.  A close inspection has revealed  that DE affects also the optical aspect which has been  dealt with  using the shadow  geometries in terms of one dimensional real curves.   For certain black holes solutions, DE can be considered as a geometric  deformation parameter controlling the   size of the shadows \cite{Belhaj:2020rdb, shad1,shad2,shad3,shad4,shad5}.\\

The aim of this work is  to  contribute to such activities by  investigating   physical behaviors of  the quintessential  Hayward-AdS black holes in four dimensions.  In particular, we generalize  certain physical properties of  the ordinary     Hayward AdS black holes without external moduli space associated with the dark sector. We elaborate a study in terms of the new quantities  $c$ and $\omega_q$ parametrizing the dark sector moduli space.   Precisely,  we  study the effect of such  parameters on certain thermodynamic and optical  aspects. Before examining the  quintessential thermodynamic  behaviors, we first  reconsider the critical properties   of the ordinary solutions associated with $c=0$.   We  find that the equation of state predicts a universal  ratio  given by $\chi_0=\frac{P_cv_c}{T_c}=\frac{27-3\sqrt{6}}{50}$.  This ration is  different than the universal  one 
found for Van der Waals fluids.   Considering  the quintessential solutions and  taking  certain values of the DE state parameter $\omega_q$,  we remark that the new ratio   depends on   the DE field intensity $c$.   In certain regions of the moduli space,  we show  that  this ratio can be factorized  using two parts   describing  the absence  and the presence of the dark sector.   By  considering  models associated with   $\omega_q=-1,-\frac{1}{3},  -\frac{2}{3}$, we  analyze  also   the DE effect on  the heat engine behaviors  of such black holes. Putting $c=0$,  we recover the previous results corresponding to the ordinary solutions.  For the optical aspect, 
 we  examine  the effect  of DE on the shadows in terms of one dimensional real curves.    We find   that  DE contributions deform the  shadow radius. Finally, we discuss the associated energy emission rate using the dark sector.

This work is organized as follows. In section 2, we reconsider  the study of the critical behaviors of  the quintessential  Hayward-AdS black holes in four dimensions.  In section 3, we investigate the effect of DE on such black holes as  heat engines. In section 4, we   examine the associated optical behaviors by considering shadow  geometries  using one dimensional real  closed curves, and  the associated energy emission.  In the last section, we give   conclusions and final remarks.

\section{ The model: Quintessential Hayward-AdS Black Hole}
It has been suggested that black holes in scalar field backgrounds could provide concrete and  semi-realistic models in connections with cosmological   findings.  An examination shows that many scalar models  have been introduced supported  by non-trivial theories   including  M-theory and superstrings.  This could produce black holes  with  external parameters  describing  the scalar field sector.  The later has been  approached  from different angles.   The most  exotic models are the dynamic dark energy models  modeled in terms of a scalar field.   It turns  out that there are several types of scalar  field  theories  including the quintessence model. It has been shown that this model can be considered as a  cooling system. This result pushes one to inspect the effect of such an energy on the other physical properties including the thermodynamical and optical ones.

The models  that   we would  like to  elaborate   are the quintessential  AdS  black holes  in four dimensions. In  the simplest  model,  the quintessential  DE could be  formulated in terms of   a  scalar field minimally coupled to gravity describing the ordinary AdS black hole solutions.  Concretely, a close  examination reveals  that the moduli space of such AdS  black holes  can 
be factorized in two sectors
\begin{equation}
\label{ }
\mathcal{M}=\mathcal{M}_{obh}\times\mathcal{M}_{ds}
\end{equation} 
The first sector  corresponds to the parameters of the ordinary AdS  balck holes (obh)
\begin{equation}
\label{ }
\mathcal{M}_{obh}=\{M,a,Q,\Lambda\}
\end{equation} 
where $a$  is the  rotating parameter and $\Lambda$ is the cosmological constant.  $M$   and $Q$ are the mass and charge parameters, respectively. For the non-rotating black hole $(a = 0)$, this can be reduced to 
a black hole  moduli space parameterized only by $M$,  $ Q$ and $\Lambda$
\begin{equation}
\label{ }
\mathcal{M}_{obh}=\{M,Q,\Lambda\}.
\end{equation} 
The second sector $\mathcal{M}_{ds}$  that we are interested in   represents the extra contributions corresponding to 
outside horizon contributions including DE, DM and other non trivial ones.
Here,  we consider only DE contributions via a quintessence scalar field. In
this way,   the dark sector    can be  controlled by two parameters $c$ and $\omega_q$     associated with 
the quintessence intensity and the DE state parameter, respectively. In this way,   we can write 
\begin{equation}
\label{ }
\mathcal{M}_{ds}=\{c,\omega_q\}.
\end{equation} 
 For generic values
of $c$,  a close examination shows that the   associated black hole  models should depend only on $\omega_q$.  In the present work, however,
we pay attention to  certain  particular cases. General values  of such a DE state  parameter may need non trivial
reflections  to analyze the corresponding  thermodynamic and optical  behaviors in the presence of   the  quintessence. This contribution  can be considered as an alternative contribution to the cosmological constant where the equation of state  could take  a central place. Many models  of such energy contributions have been dealt  with  including ones describing    a possible   deviation of the cosmological constant  via  a single scalar field.  In this way, the equation of state parameter can be toke as  the following form
\begin{eqnarray}
\omega_q= -1+\frac{2}{3}\epsilon_\phi
\end{eqnarray}  
where  $\epsilon_\phi $  indicates   a positive     contribution obtained from   the  scalar potential associated with   the   quintessence. In cosmological models,   $\epsilon_\phi$  has been called slow-roll parameter. 
  Certain  models could be  examined    using  scalar potentials
already investigated  in the literature.  The selection of such scalar models  has  not  been an easy task.  In connection with  black holes,  three models  have been extensively studied corresponding to   three values of  $ \epsilon_\phi$ being  0, $ \frac{1}{2}$, and 1 giving    $\omega_q=-1,-\frac{2}{3},  -\frac{1}{3}$.   Assuming that  the spacetime metric  is  static, spheric and symmetric,  the  line element   describing such non rotating AdS black holes should be written as 
\begin{equation}
ds^2=-f_{\omega_q}(r)dt^2+\frac{dr^2}{f_{\omega_q}(r)}+r^2d\Omega^2,
\label{1254}
\end{equation}
where   $f_{\omega_q}$   is the  quintessential  black hole  metric function and $d\Omega^2$ is the line element of  the 2-dimensional unit sphere.   This function, which  contains information on the above factorized moduli space, can be written as 
\begin{equation}
f_{\omega_q}(r)=f(r)+cg(r)
\label{1254}
\end{equation}
where  $g(r)$  is a function  which depends  on   the dark sector state equation.  Taking $c=0$,  we can  recover the ordinary  the  AdS Black Holes described by the  metric function $f$.
 The main objective  of this paper is to inspect  the  influence of the   dark sector   moduli space  $\mathcal{M}_{ds}$ on  physical properties of  the  ordinary Hayward  AdS black holes.   
In particular, we study DE effects  in terms of the new parameters  associated with   such a sector.  At particular points of  the moduli spaces, we will show that   certain   thermodynamical quantities $X_c$  can be factorized   according the above metric function as follows
\begin{equation}
X_c\sim X_0+cX
\label{1254}
\end{equation}
where $X$ is an extra contribution associated with  the dark sector.  Similar optical  behaviors will be  shown. 
\section{Thermodynamic  behaviors of Hayward-AdS black hole   from DE}
In this section, we   investigate the effect of DE  on critical behaviors of  the   Hayward-AdS black holes in four dimensions.  Before studying such models, we reconsider the study of the ordinary solutions without DE contributions. In particular, we investigate the critical aspect.

\subsection{Criticality behaviors  of ordinary black hole solutions}
\begin{figure}[ht!]
		\begin{center}
		\centering
			\includegraphics[scale=.6]{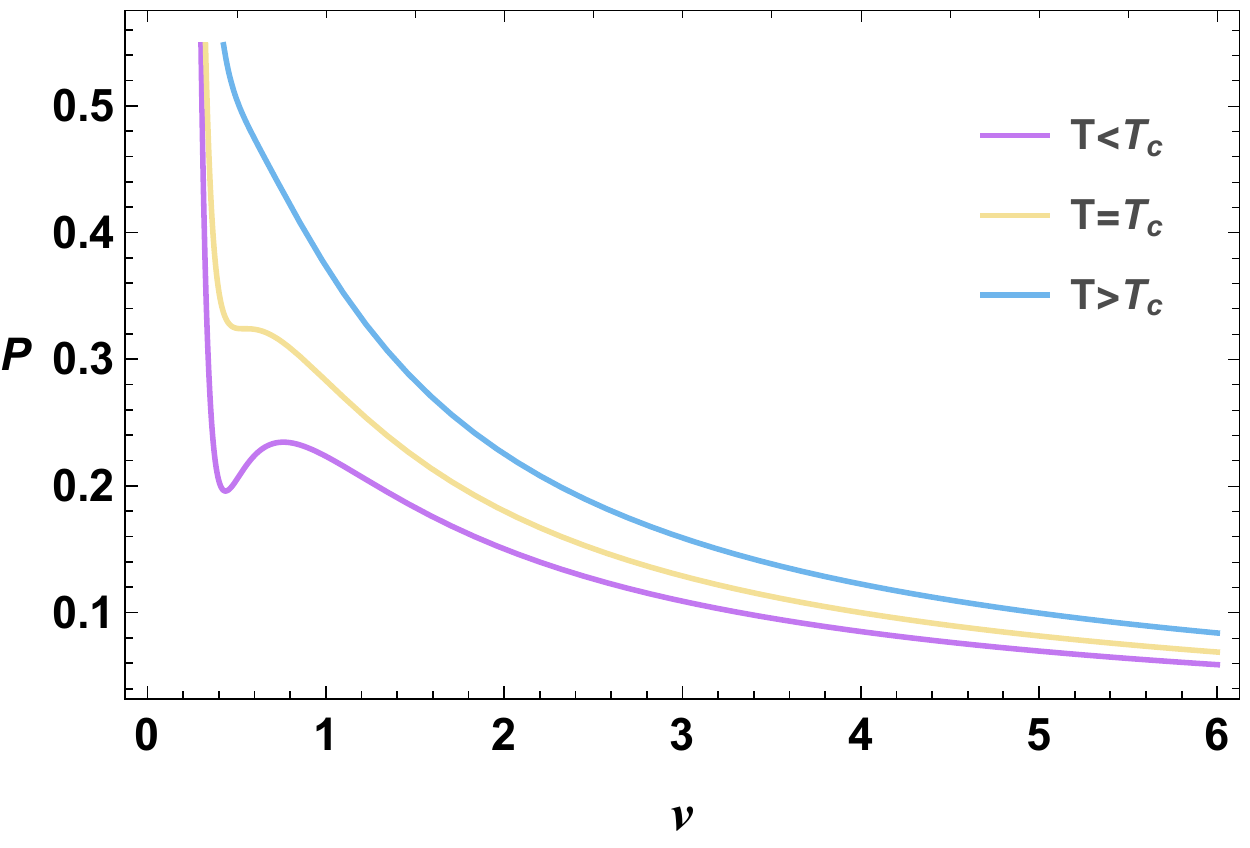}
\caption{{\it \footnotesize$P-v$ digram of the Hayward-AdS black hole for different values   of $T$ by taking $Q=0.1$.}}
\label{f0}
\end{center}
\end{figure}

In this part, we investigate the criticality behaviors of the proposed  black hole solutions without DE contributions.  For generic charge values,  we first  show that the ordinary solutions involve a  critical universal ratio. Precisely,   the  equation of state for  the quintessential black holes  predicts a critical universal  ratio  depending on  the involved DE fields.   To reveal that,   we   consider   the ordinary solutions associated  with  the  Hayward-AdS metric \cite{31}. 
We start by  taking  the following action
 \begin{equation}
S=\frac{1}{16\pi}\int d^4x\sqrt{-g}\left(R-2\Lambda-\mathcal{L}(\mathcal{F})\right),
\end{equation}
where  one has used $\mathcal{F}=F_{\mu\nu}F^{\mu\nu}$. Here,   $\mathcal{L}$ is a Lagrangian density depending only on $\mathcal{F}$ and $\Lambda=-\frac{3}{\ell^{2}}$. Varying the action with respect to the metric tensor $g_{\mu\nu}$ and  the field strength $F_{\mu\nu}$, one gets  the following  equations of motion
\begin{eqnarray}
G_{\mu\nu}-\frac{3}{\ell^2}g_{\mu\nu}=T_{\mu\nu}, \quad  \quad 
\nabla_\mu\left(\mathcal{L}_{\mathcal{F}}F^{\mu\nu}\right)=0
\end{eqnarray}
with  $\mathcal{L}_{\mathcal{F}}=\frac{\partial\mathcal{L}(\mathcal{F})}{\partial \mathcal{F}}$ and $
T_{\mu\nu}=2\left(\mathcal{L}_{\mathcal{F}}F_{\mu\nu}^2-\frac{1}{4}g_{\mu\nu}\mathcal{L}\right).$
For $c=0$,   the spacetime metric   has  the following line element  
\begin{equation}
ds^2=-f(r)dt^2+\frac{dr^2}{f(r)}+r^2d\Omega^2.
\label{1254}
\end{equation}
 Using the above equations, we can obtain the $(t,t)$ component  metric  giving 
\begin{equation}
\frac{rf'(r)+f(r)-1}{r^2}-\frac{3}{\ell^2}=-\frac{\mathcal{L}(\mathcal{F})}{2}.
\end{equation} 
In terms of  the mass parameter,  this can be written as 
\begin{equation}
\frac{2m'(r)}{r^2}+\frac{3}{\ell^2}=\frac{\mathcal{L}(\mathcal{F})}{2}.
\end{equation}
The associated calculation can be carried out according to the work reported in \cite{ddd}.  In particular,  we consider  the Lagrangian density  given by
\begin{equation}
\mathcal{L}(\mathcal{F})=\frac{4\mu}{\alpha}\frac{\left(\alpha\mathcal{F}\right)^{5/4}}{\left(1+\sqrt{\alpha\mathcal{F}}\right)^{1+\frac{\mu}{2}}}
\end{equation}
where one has $\mathcal{F}=\frac{2q_m^2}{r^4}$.   $\alpha$ is a constant satisfying  $\alpha=\frac{Q^4}{2q_m^2}$.  $\mu$ is a positive  dimensionless constant. It is recalled that Hayward black hole can be   recovered  by taking  $\mu=3$ and $\alpha^{-1}Q^3=M$. After certain calculations,  one can obtain 
\begin{equation}
m(r)=-\frac{8Mq_m^6}{8q_m^6+M^3r^3}+C_1-\frac{r^3}{2\ell^2},
\end{equation}
where $C_1$ is an integration constant and  where  $M$ is the  black hole  mass parameter.   Many solutions could  be obtained by taking certain limits. Here,  we consider   $f(r)$   which is a relevant  function known by the  metric function    taking the following form
\begin{equation}
f(r)=1+\frac{r^{2}}{\ell^{2}}-\frac{2Mr^{2}}{r^{3}+Q^{3}},
\end{equation}
where $\ell$  represents the AdS length.  It is  noted that the parameter $Q$ is related to the total magnetic charge $q_{m}$   via the relation
\begin{equation}
q_{m}=\frac{Q^{2}}{\sqrt{2a}}
\end{equation}
where $a$ is a free integration constant. In thermodynamical activities of the  AdS black holes,  the  cosmological constant $\Lambda = -\frac{3}{\ell^{2}}$  has been  interpreted  as the  pressure
\begin{equation}
P = -\frac{\Lambda}{8 \pi} = \frac{3}{8 \pi \ell^{2}},
\end{equation}
and its conjugate variable is associated with the   thermodynamic volume.
The   black hole mass  $M$ can be  obtained by solving the constraint $f(r)= 0$.  The computations give 
\begin{equation}
M=\frac{\left(Q^3+r_h^3\right) \left(8\,P\,\pi\,r_h^2+3\right)}{6\,  r_h^{2}}.
\label{1a}
\end{equation}
To  obtain  the associated temperature, the first law of the black hole thermodynamics  is needed. It is formulated by 
\begin{equation}
dM = TdS + VdP + \phi dq_m+\Pi da,
\end{equation}
where  $T$, $\phi$,  $V$  and $\Pi$ are the temperature, the  electrostatic potential,  the volume thermodynamic and the quantity conjugate  to $a$,
respectively \cite{Fan}. After calculations, we find the following   temperature  
\begin{equation}
T=\frac{\kappa}{2\pi}=\frac{1}{4}\frac{\partial f(r) }{\partial r} \bigg\vert_{r=r_h}=\frac{8 \pi  P r_h^5-2 Q^3+r_h^3}{4 \pi  r_h \left(Q^3+r_h^3\right)}.
\label{temp1}
\end{equation}
The  entropy of such a  black hole  solution   is given by 
\begin{equation}
S=\int{\frac{dM}{T}}=\pi r_h^2.
\label{2}
\end{equation}
Exploiting  Eq.(\ref{1a}) and Eq.(\ref{2}),  the computations provide the following   generalized mass formula
\begin{equation}
M(S,Q,P)=\frac{(\pi^{\frac{3}{2}}\,Q^{3}+S^{\frac{3}{2}})(3+8\,P\,S)}{6\,\sqrt{\pi}\,S}.
\end{equation}
According  to the method explored in  many places,  one can obtain  the equation of state for such  four-dimensional    AdS-black holes \cite{28,29,30}. For a generic point in the   associated   reduced  moduli space,  we obtain 
\begin{equation}
P=\frac{16 Q^3 (\pi  T v+1)+v^3 (2 \pi  T v-1)}{2 \pi  v^5},
\label{3a}
\end{equation}
where $v$  represents  the specific volume  given by
\begin{equation}
v=2r_h.\end{equation}
Having obtained   the relevant thermodynamical  quantities, we move to   analyze  the associated  $P-v$  diagram. Indeed, the corresponding behaviors  are plotted in the Fig.(\ref{f0}).
It has been observed from this figure that for the values of the temperature higher to the critical one $T_C$, the system behaves  like an  ideal gas. When  the  temperature  takes the value $T_C$, one could talk about   the  the critical isotherm being characterized by an inflection point corresponding to the critical pressure $P_C$ and the critical volume $v_C$. The values of  the temperature lower to $T_C$  represent the unstable thermodynamic region.   The critical points 
should  verify  the  following constraints
\begin{equation}
\frac{\partial P}{\partial \upsilon}=0,   \hspace{1.5cm}
\frac{\partial^{2}P}{\partial \upsilon^{2}}=0.
\end{equation}
After computations,  we find the critical quantities 
\begin{eqnarray}
T_C & = & \frac{\left(5 \sqrt{2}-4 \sqrt{3}\right)\left(7+3 \sqrt{6}\right)^{2/3}}{4\ 2^{5/6}\, \pi \, Q}, \\
v_C & = &{16}^{1/3}  \left(7 Q^3+{54}^{1/2} Q^3\right)^{1/3},\\
P_C & = & \frac{3 \left(\sqrt{6}+3\right)}{16\ 2^{2/3} \left(3 \sqrt{6}+7\right)^{5/3} \pi  Q^2}.
\end{eqnarray}
For the reduced moduli space associated with generic charges values, these critical values provide   a critical universal  ratio
\begin{eqnarray}
\chi_0=\frac{P_Cv_C}{T_C}=\frac{27-3\sqrt{6}}{50}\simeq0.393.
\label{bv}
\end{eqnarray}
A this level, one can provide certain comments on such a ratio.  First,  it  is noted  that the temperature can be obtained using an alternative way by varying the mass with respect to the entropy.  Such a way provides a closed number given by $\frac{2}{5}$.  Second, it  is observed  that this  ratio  number is different than  the one obtained in   the  charged AdS black holes being  $\frac{P_Cv_C}{T_C}=3/8$ found also  in  Van der Waals fluids \cite{Kubiznak:2012wp}.  Finally,  a close examination shows that  the function metric of  the Hayward-AdS black hole  is the relevant  responsible of  such a distinction.
\subsection{Criticality of  quintessential solutions}
\begin{figure}[ht!]
		\begin{center}
		\centering
			\begin{tabbing}
			\centering
			\hspace{8.2cm}\=\kill
			\includegraphics[scale=.55]{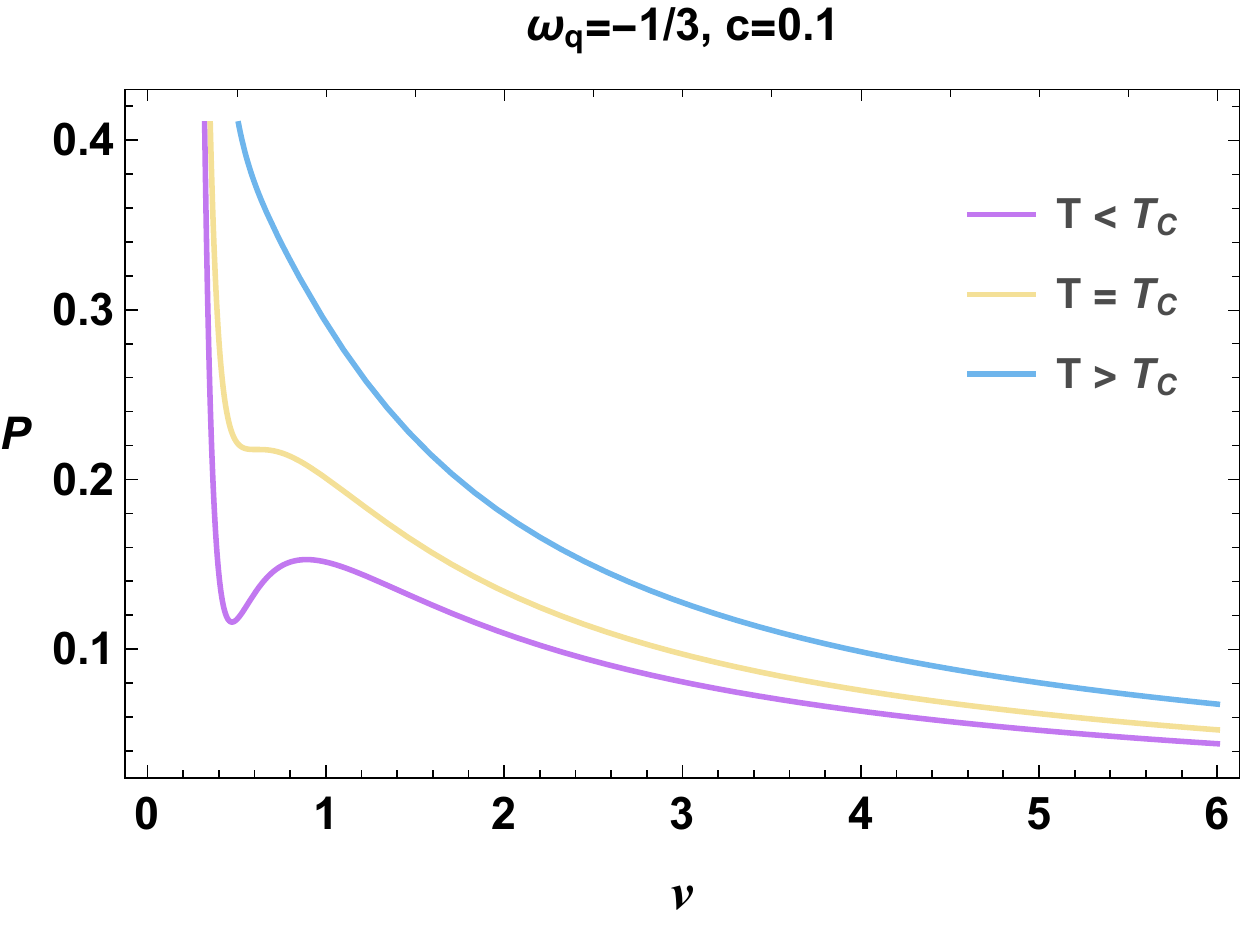} \>
			\includegraphics[scale=.55]{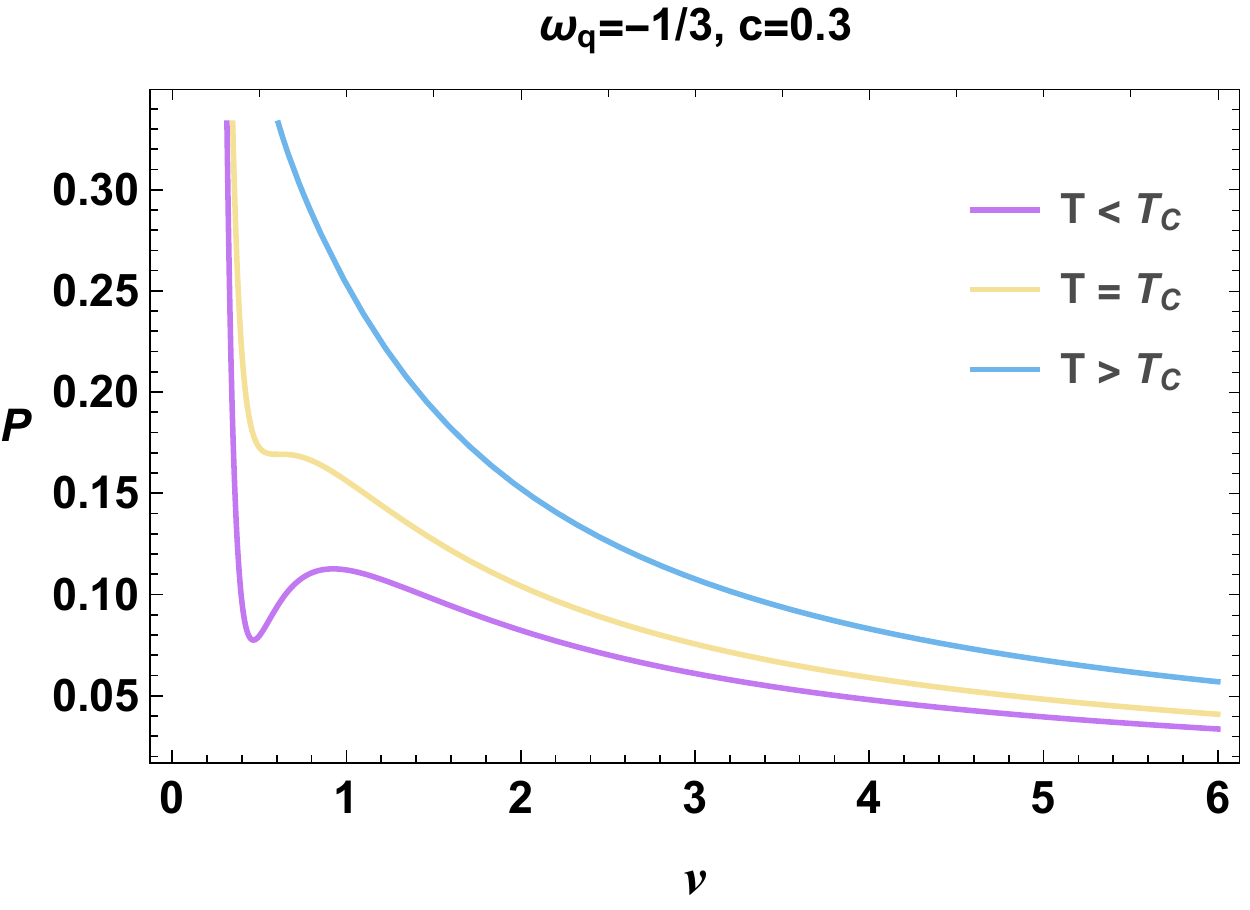} \\
			\includegraphics[scale=.55]{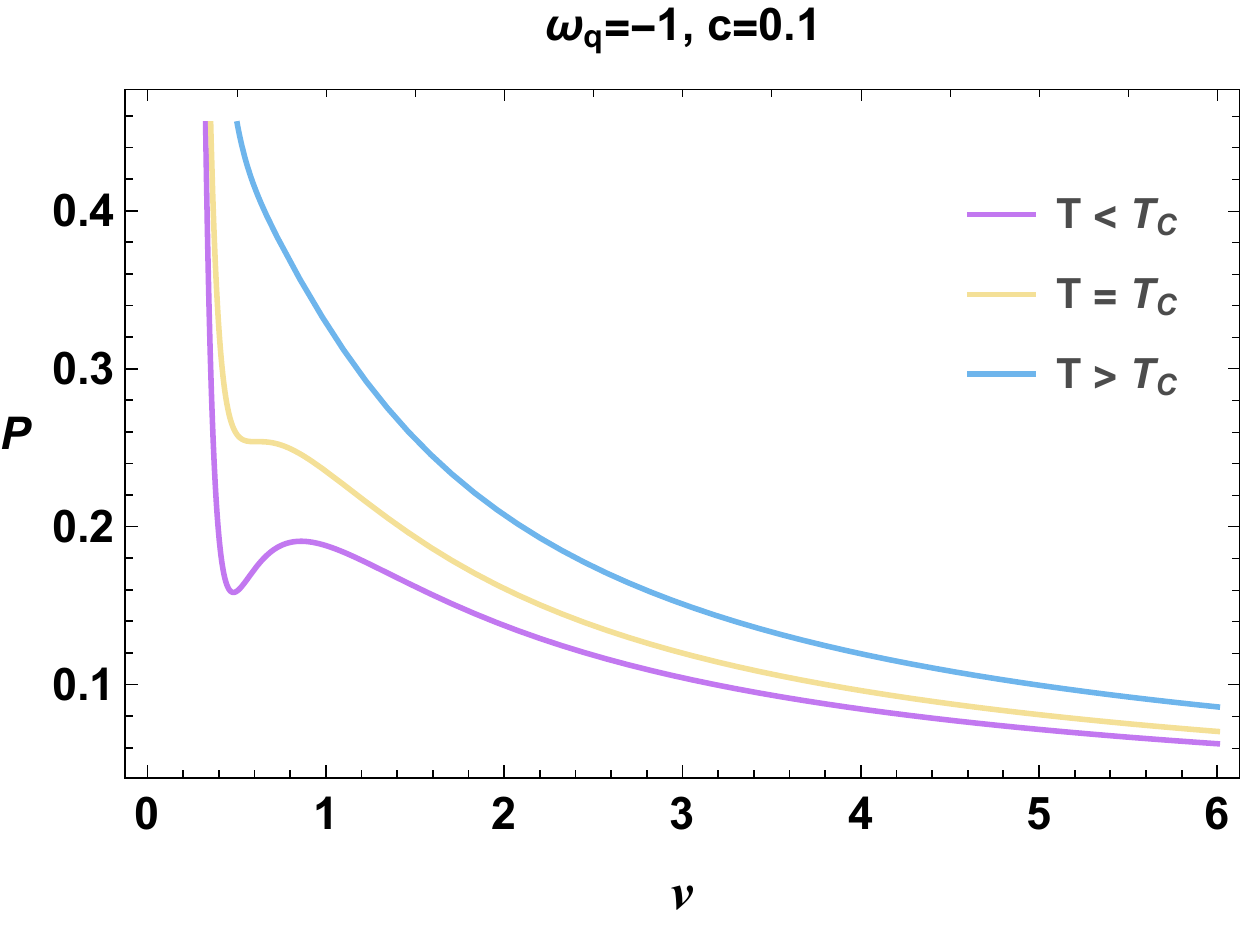} \>
			\includegraphics[scale=.55]{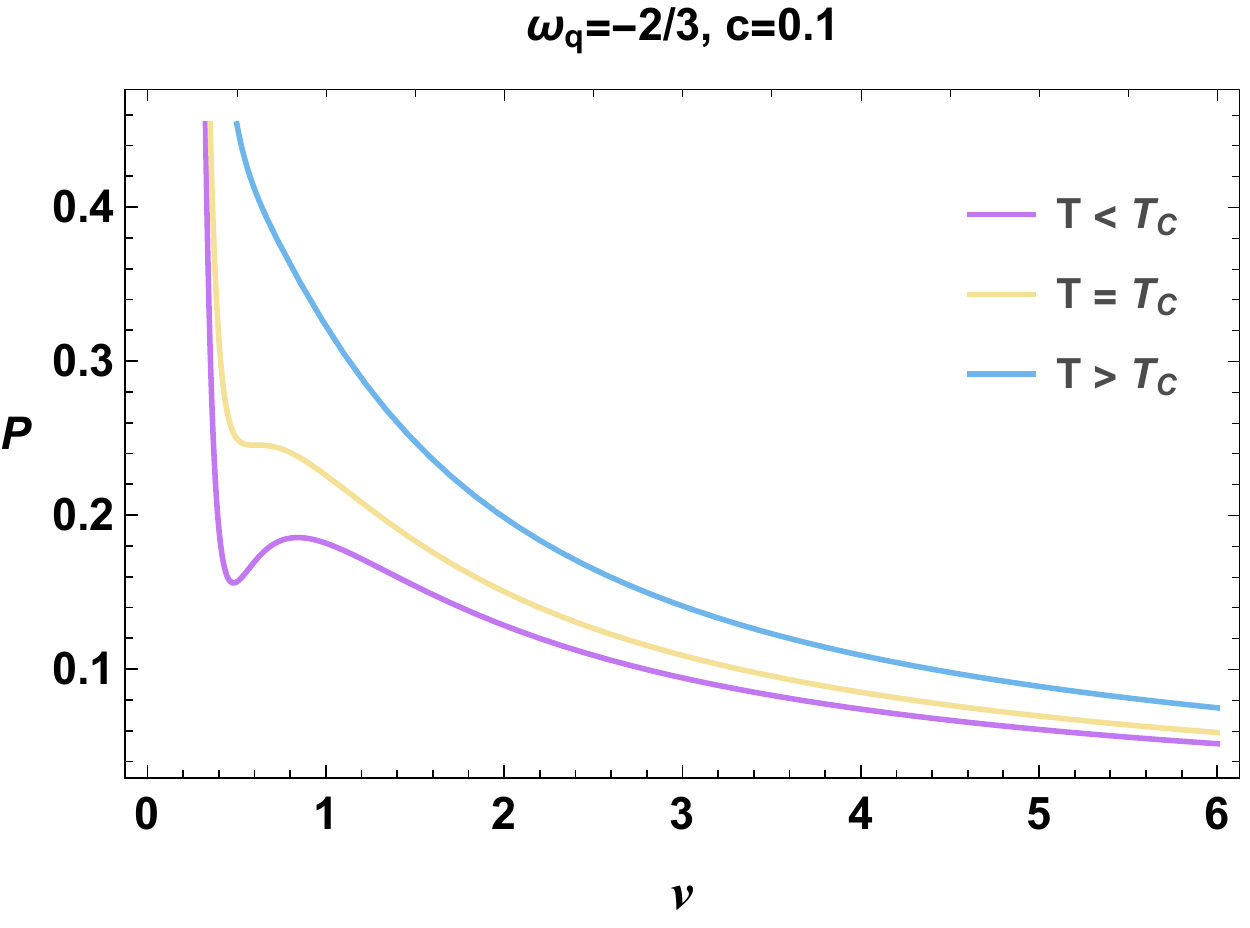} \\
		   \end{tabbing}
		   \vspace{-1cm}	
\caption{{{\it \footnotesize$P-v$ diagram for Hayward-AdS black hole for different values of  $T$, $c$, and $\omega_q$  by taking $Q=0.1$. }}}
\label{ex1}
\end{center}
\end{figure}

Now  we are in position to   consider   the DE  effect  on such critical behaviors by introducing a quintessence scalar  field. In this  regard, the  study will be made    in terms of  two parameters $c$ and  $\omega_q$  associated with the quintessence intensity and the DE state parameter, respectively.   Concretely, we will be interested in how the obtained critical universal number  behaves  in terms of such a DE field  related to the density of quintessence  via the relation 
\begin{equation}
\rho_{q}=-\frac{c}{2}\frac{3\omega_{q}}{r^{3(\omega_{q}+1)}}.
\label{111}
\end{equation}
 It turns out that one can anticipate a relevant  behavior for  small values of $c$.   To visualize  the effect of DE in the critical thermodynamic  quantities,  we  propose the following  relation  
\begin{equation}
\chi \sim \chi_0+\xi_{\omega_q}(c,Q).
\end{equation}
  This  relation separates the ordinary contributions of the  Hayward-AdS black holes and the ones  of  DE. In  fact,  $\xi_{\omega_q}$ represents  the DE contributions. However,  $\chi_{0}$  denotes  the contribution without DE, given by the Eq (\ref{bv}).  According to  \cite{Kiselev:2002dx}, and  taking   the Einstein equations for static black holes surrounded by the  quintessence where   the stress-energy tensor  involves the additivity and linearity conditions, one finds
\begin{eqnarray}
T^t_t & = & T^r_r = \rho_q, \\
T^\theta_\theta & = & T^\phi_\phi = -\frac{\rho_q}{2}(3\omega_q+1).
\end{eqnarray}
In this way, the presence of  the quintessence DE field requires that the above metric function $f(r)$ of the Hayward-AdS black hole should be modified as follows
\begin{equation}
f_{\omega_q}(r)=1+\frac{r^{2}}{\ell^{2}}-\frac{2Mr^{2}}{r^{3}+Q^{3}}-\frac{c}{r^{3\omega_{q}+1}},
\end{equation}
where $f_{\omega_q}$  is the metric function in the presence of DE.  Using   Eq.(\ref{1254}), we get  an exact solution of the Hayward AdS black hole in presence of the quintessential field\cite{ddd}. As before, the black hole mass $M$ can be  obtained by solving the constraint $f_{\omega_q}(r)= 0$.  Precisely,  we  find
\begin{equation}
M=\frac{\left(Q^3+r_h^3\right) \left((8\,P\,\pi\,+1)r_h^{3 \omega_q +1} -3\,c \right)}{6\,  r_h^{3 \omega_q +3}}.
\label{1}
\end{equation}
It is remarked that the previous mass equation can be recovered by sending $c$ to zero. Exploiting  Eq.(\ref{1}) and Eq.(\ref{temp1}) and using  the first law of the black hole thermodynamics, the   temperature  is found to be
\begin{equation}
T(\omega_q,c)=\frac{1}{4}\frac{\partial f_{\omega_q} }{\partial r} \bigg\vert_{r=r_h}=\frac{8 \pi  P r_h^5-2 Q^3+r_h^3}{4 \pi  r_h \left(Q^3+r_h^3\right)}-\frac{3  r^{-(3 \omega_q +1)}_h \left(Q^3 (\omega_q +1)+r_h^3 \omega_q \right)c}{{4 \pi  r_h \left(Q^3+r_h^3\right)}.}.
\label{temp}
\end{equation}
According  to the method explored in \cite{Kubiznak:2012wp},  one can obtain  the equation of state for such  four-dimensional  modified   AdS-black hole solutions. For a generic point in the   associated   moduli space, the computations  provide the following state equation
\begin{equation}
P(\omega,c)=\frac{16 Q^3 (\pi  T v+1)+v^3 (2 \pi  T v-1)}{2 \pi  v^5}-\frac{3 \times  2^{3 \omega_q +1} v^{-(3 \omega_q+1) } \left(8 Q^3 (\omega_q +1)+v^3 \omega_q \right)c}{2 \pi  v^5}
\label{6}
\end{equation}
where $v$  represents  the associated  specific volume. Taking $c=0$,  we recover the ordinary pressure  appearing in Eq.(\ref{3a}). \\ Having obtained the relevant thermodynamical  quantities, we will be  interested in analytical and numerical analysis on the obtained results. Precisely, we discuss the $P-V$ diagram.   Motivated by similar activities, we  deal with specific values of $\omega_q$ and $c$ being extensively studied  in connections with  the DE   contributions. Indeed,   the corresponding behaviors   are  plotted in  Fig.(\ref{ex1}).
It has been observed that  for different values of DE parameters, we  find  similar behaviors  for  the    $P-v$ diagram   appearing in  the ordinary  Hayward-AdS black hole solutions without DE contributions.  However,  the  one difference is  that the unstable thermodynamic region associated with  such values  of  the DE parameter is relevant with respect to   the  ordinary ones.   Taking  different values of $\omega_q$ and  considering  $c=0.1$, the critical  quantities  are  almost the same. For $\omega_q=-1/3$ and $c=0.3$, however, the critical  values   are lower    with respect to  other models.   Similar behaviors appear in   the  ordinary solutions.   In the $P-V$ diagram,  the inflection points should   verify   the following constraints 
\begin{equation}
\frac{\partial P(\omega,c)}{\partial \upsilon}=0,   \hspace{1.5cm}
\frac{\partial^{2}P(\omega,c)}{\partial \upsilon^{2}}=0.
\label{111}
\end{equation}
 An examination reveals that the solution of equation (\ref{111})  can be obtained by taking  a fixed  value of $\omega_{q}$. This will be exploited to  get the expression of  the critical  quantities  $T_{c}$, $v_{c}$,  and $P_{c}$.
Instead of   taking generic expressions,  we  examine  only three different $\omega$-models.
For $\omega_q$=-1, we obtain the following critical quantities 
\begin{eqnarray}
T_C & =&   \frac{\left(5 \sqrt{2}-4 \sqrt{3}\right)\left(7+3 \sqrt{6}\right)^{2/3}}{4\ 2^{5/6}\, \pi \, Q} \\
v_C & = &{16}^{1/3}  \left(7 Q^3+{54}^{1/2} Q^3\right)^{1/3}\\
P_C & = & \frac{3 \left(\sqrt{6}+3\right)}{16\ 2^{2/3} \left(3 \sqrt{6}+7\right)^{5/3} \pi  Q^2}+\frac{c}{8\pi}.
\end{eqnarray}
In  this model, the  behavior of the extended universal ratio  takes a  nice form. Precisely,  it  can be written 
\begin{equation}
\chi_c = \chi_0+\xi_{-1}(c,Q).
\end{equation}
After  calculations, we find 
\begin{equation}
\xi_{-1}(c,Q)=\frac{3\ 2^{2/3} \left(416 \sqrt{6}+1019\right) c Q^2}{\left(3 \sqrt{6}+7\right)^{7/3}}.
\label{x1}
\end{equation}
This is a nice general expression since it gives, as a particular case for $c =0$,   the value  $\frac{27-3\sqrt{6}}{50}$ for generic charge values. Taking  $\omega_q=-\frac{2}{3}$, due to the higher degree of the critical parameter  equation, we can prove  numerically for different values of $Q$ and $c$ that the critical  quantities  can produce a critical universal number. Indeed, it   can be written as 
\begin{equation}
\chi_c \sim \chi_0+\xi_{-\frac{2}{3}}(c,Q).
\end{equation}
The associated calculations are given in  table.(\ref{T13}). For  a vanishing value of the $c$ parameter, this  can be reduced to the usual value $\chi=\frac{27-3\sqrt{6}}{50}=0.393$ appearing in the black holes  without DE contributions given in Eq.(\ref{bv}).
Taking  $\omega_q=-\frac{1}{3}$, we get
\begin{eqnarray}
T_C & =&   \frac{\left(5 \sqrt{2}-4 \sqrt{3}\right)\left(7+3 \sqrt{6}\right)^{2/3}(1-c)}{4\ 2^{5/6}\, \pi \, Q}, \\
v_C & = &{16}^{1/3}  \left(7 Q^3+{54}^{1/2} Q^3\right)^{1/3},\\
P_C & = &\frac{3 \left(\sqrt{6}+3\right) (1-c)}{16\ 2^{2/3} \left(3 \sqrt{6}+7\right)^{5/3} \pi  Q^2}.
\end{eqnarray}
This provides exactly Eq.(\ref{bv}).  For  $c=0$,  we can  show that the critical thermodynamic  quantities   are  equivalents  to the  Hayward black hole solutions in EGB gravity by taking the limit $\alpha\to 0$\cite{28}. In this model,  the  expression of the universal ratio  does not   depend on the $(Q, c)$ moduli space. It is valid for generic  charge values of  such  quintessential  black holes.\\
To  inspect  the influence $\omega_q$   for  generic regions of the $(Q, c)$ moduli space,  we  should compute  the ratio    $\frac{P_{C}v_{C}}{T_{C}}$.    The calculations  are  listed in Table.(\ref{T13}).
\begin{table}[ht!]
\begin{center}
\scalebox{0.72}{
\begin{tabular}{|l|c|c|c|c|c|c|c|c|c|c|c|}
\hline
\multicolumn{2}{|l|}{\cellcolor[HTML]{EFEFEF}{\color[HTML]{343434} }}                   & \cellcolor[HTML]{EFEFEF}            & \multicolumn{3}{c|}{$\omega=-1$}              & \multicolumn{3}{c|}{$\omega=-\frac{2}{3}$} & \multicolumn{3}{c|}{$\omega=-\frac{1}{3}$} \\ \cline{3-12}
\multicolumn{2}{|l|}{\multirow{-2}{*}{\cellcolor[HTML]{EFEFEF}{\color[HTML]{343434} }}} & \multicolumn{1}{l|}{\textbf{$c=0$}} & \textbf{$c=0.1$} & \textbf{$c=0.2$} & $c=0.3$ & $c=0.1$      & $c=0.2$      & $c=0.3$      & $c=0.1$      & $c=0.2$      & $c=0.3$      \\ \hline
\multicolumn{1}{|c|}{}                                   & $T_c$                        & 0.376                               & 0.376            & 0.376            & 0.376   & 0.363        &0.350         &0.337          & 0.339        & 0.301        & 0.263        \\ \cline{2-12}
\multicolumn{1}{|c|}{}                                   & $v_c$                        & 0.612                               & 0.612           & 0.612            & 0.612   &0.608        &  0.604       &   0.600     &  0.612         & 0.612        & 0.612        \\ \cline{2-12}
\multicolumn{1}{|c|}{}                                   & $P_c$                        & 0.241                               & 0.265            & 0.347            & 0.277   & 0.245       & 0.249        & 0.252        &  0.217        & 0.193     & 0.169      \\ \cline{2-12}
\multicolumn{1}{|c|}{\multirow{-4}{*}{$Q=0.1$}}          & $ {\bf\frac{P_cv_c}{T_c}}$         &  {\bf0.393}                                 &  {\bf0.412}            &  {\bf0.431}            &  {\bf0.451}   &  {\bf0.410}        &  {\bf0.429}        &  {\bf0.449}        &  {\bf0.393}        &  {\bf0.393}        &  {\bf0.393}        \\ \hline
\multicolumn{1}{|c|}{}                                   & $T_c$                        & 0.188                               & 0.188            &  0.188             &  0.188    & 0.175        & 0.162        & 0.149        & 0.169        & 0.150        & 0.131        \\ \cline{2-12}
\multicolumn{1}{|c|}{}                                   & $v_c$                        & 1.224                               & 1.224            & 1.224            & 1.224  & 1.209        & 1.194        & 1.178        & 1.224        & 1.224        & 1.224      \\ \cline{2-12}
\multicolumn{1}{|c|}{}                                   & $P_c$                        & 0.060                               & 0.072            & 0.084           & 0.096   & 0.062        & 0.075        & 0.088        & 0.054        & 0.048        & 0.042        \\ \cline{2-12}
\multicolumn{1}{|c|}{\multirow{-4}{*}{$Q=0.2$}}          & $ {\bf\frac{P_cv_c}{T_c}}$         &  {\bf0.393}                                 &  {\bf0.470}            &  {\bf0.548}            &  {\bf0.625}   &  {\bf0.429}        &  {\bf0.552}        &  {\bf0.693}        &   {\bf0.393}        &   {\bf0.393}        &   {\bf0.393}        \\ \hline
                                                         & $T_c$                        & 0.125                               & 0.125            & 0.125            & 0.125   & 0.112        & 0.099        & 0.087        & 0.113        & 0.100        & 0.087        \\ \cline{2-12}
                                                         & $v_c$                        & 1.836                               & 1.836            & 1.836            & 1.836   & 1.802        & 1.768        & 1.734        & 1.836        & 1.836        & 1.836        \\ \cline{2-12}
                                                         & $P_c$                        & 0.026                               & 0.038            & 0.050            & 0.062   & 0.092        & 0.105        & 0.117        & 0.024        & 0.021        & 0.018       \\ \cline{2-12}
\multirow{-4}{*}{$Q=0.3$}                              & $ {\bf\frac{P_cv_c}{T_c}}$         & {\bf0.393}                                 & {\bf0.567}            & {\bf0.742}            & {\bf0.916}   & {\bf1.486}        & {\bf1.863}        & {\bf2.336}        & {\bf0.393}        & {\bf0.393}        & {\bf0.393}       \\ \hline
                                                         & $T_c$                        & 0.094                               & 0.094            & 0.094            &  0.094   & 0.095        & 0.081        & 0.067        & 0.084        & 0.075        & 0.065        \\ \cline{2-12}
                                                         & $v_c$                        & 2.449                               & 2.449            &2.449            & 2.449   & 2.147        & 2.124        & 2.100        & 2.449        & 2.449       & 2.449        \\ \cline{2-12}
                                                         & $P_c$                        & 0.015                               & 0.038            & 0.044            & 0.050   & 0.020        & 0.021        & 0.022        & 0.013        & 0.012       & 0.010        \\ \cline{2-12}
\multirow{-4}{*}{$Q=0.4$}                                & $ {\bf\frac{P_cv_c}{T_c}}$         &  {\bf0.393}                                 &  {\bf0.703}            &  {\bf1.013}            &  {\bf1.324}   &  {\bf0.467}        &  {\bf0.558}        &  {\bf0.688}        & {\bf0.393}        &{\bf0.393}        & {\bf0.393}        \\ \hline
\end{tabular}}
\caption{{\it \footnotesize Numerical values of critical thermodynamic parameters in the presence of DE.  }}
\label{T13}
\end{center}
\end{table}
It has been observed that the  expression of the critical  universal  ratio   has certain  nice features. For $\omega_q$ models,  it  depends on the $(Q, c)$ moduli space.  Fixing the value of $\omega_q$  and considering  the line $(Q,0)$ of the reduced moduli space, it reduces to the usual value   $\chi_0=0.393$. At  a  generic point of the moduli space, we  remark that when $\omega_q$
increases, the quantity $\frac{P_{c}\nu_{c}}{T_{c}}$ decreases by approaching the value  $\chi_0$.   Taking  $\omega_q=-\frac{1}{3}$, in generic regions of the moduli space,   it has  been  observed  that this  critical ratio is  independent of  the DE parameter for all range of $c$.  This can be understood from the fact that the associated model could correspond to a possible length  scaling for $\omega_{q}=-\frac{1}{3}$
\begin{equation}
 \frac{1}{\ell^{2}}  \to \frac{1}{\ell^{2}}-c.
\end{equation}
For the ordinary solution and  the $\omega_q=-\frac{1}{3}$ model, we get the same value of  the ratio  $\frac{P_{C}\nu_{C}}{T_{C}}$. This critical ratio  is closed to  the value $\frac{3}{8}$,  obtained  in  AdS  black hole solutions \cite{thermo1,thermo2}.  However, for $\omega_{q}=-\frac{2}{3}$ and $\omega_{q}=-1$ models, the ratio  $\frac{P_{c}\nu_{c}}{T_{c}}$ increases  with $c$ and $Q$. Taking  small values of $c$ and $Q$, this ratio   increases slightly.  This shows  that the effect of $c$ is negligible for  such  models.  
\section{Heat engine behaviors from  quintessence}
\begin{figure}[!ht]
\begin{minipage}{\linewidth}
      \centering
      \begin{minipage}{0.45\linewidth}
          \begin{tikzpicture}
[
    corner/.style={
        circle,
        fill,
        inner sep=1.5pt},
    arrowline/.style={
        thick,
        postaction = decorate,
        decoration = {markings,
            mark = at position .8 with \arrow{latex}
        }
    },
    every pin/.style = {
        black!80,
        inner sep = 1mm,
        align = center,
        font = \footnotesize,
        pin edge = {-latex, thin, line to}},
]

\draw [latex-latex] (0,4.5)  |- (6.5,0)
    node [pos=1.02,yshift=0.02cm] {$V$}
    node [pos=0.07,yshift=0.9cm]  {$P$};

\node[corner,
    label = {left:$1$}] (1) at (1,4){} ;

\node[corner,
    label = {right:$2$}] (2) at (5.5,3){} ;

\node[corner,
    label = {right:$3$}] (3) at (5.5,0.5){} ;

\node[corner,
    label = {below left:$4$}] (4) at (1,1.5){} ;

\draw [arrowline,very thick,green] (1) to [bend right = 15]
    node [pos = .5,yshift=0.6cm] {$T_H$} (2);

\draw [arrowline,red,very thick ] (2) to []
    (3);

\draw [arrowline,green,very thick] (3) to [bend left  = 15]
    node [pos = .5,yshift=-0.3cm] {$T_C$} (4);

\draw [arrowline,red,very thick ] (4) to []
     (1);

\end{tikzpicture}
      \end{minipage}
      \hspace{0.07\linewidth}
      \begin{minipage}{0.45\linewidth}
          \begin{tikzpicture}
[
    corner/.style={
        circle,
        fill,
        inner sep=1.5pt},
    arrowline/.style={
        thick,
        postaction = decorate,
        decoration = {markings,
            mark = at position .8 with \arrow{latex}
        }
    },
    every pin/.style = {
        black!80,
        inner sep = 1mm,
        align = center,
        font = \footnotesize,
        pin edge = {-latex, thin, line to}},
]

\draw [latex-latex] (0,4.5)  |- (6.5,0)
    node [pos=1.02,yshift=0.02cm] {$V$}
    node [pos=0.07,yshift=0.9cm]  {$P$};

\node[corner,
    label = {left:$1$}] (1) at (1,3){} ;

\node[corner,
    label = {right:$2$}] (2) at (5.5,3){} ;

\node[corner,
    label = {right:$3$}] (3) at (5.5,0.5){} ;

\node[corner,
    label = { left:$4$}] (4) at (1,0.5){} ;

\draw [arrowline,very thick,green] (1) to []
    node [] {} (2);

\draw [arrowline,red,very thick ] (2) to []
    (3);

\draw [arrowline,green,very thick] (3) to []
    node [] {} (4);

\draw [arrowline,red,very thick ] (4) to []
     (1);

\end{tikzpicture}
      \end{minipage}
  \end{minipage}
  \caption{\footnotesize {\bf Left: } Carnot cycle. {\bf Right: } The studied cycle.}\label{cycleF}
  \label{fz}
\end{figure}
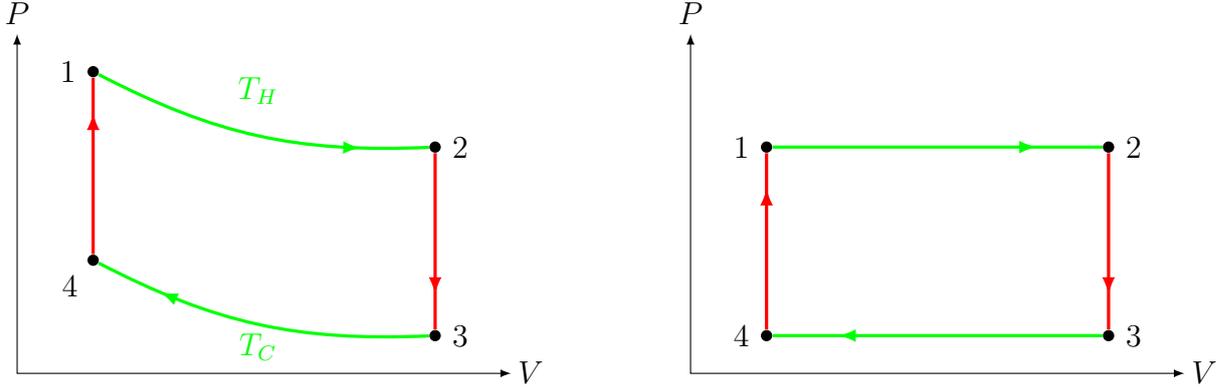

  Considering  the Hayward-AdS black holes   as  heat engines, we  compute  and investigate the associated  efficiency.  In particular, we   inspect  the effect of the quintessence field on such heat engine behaviors.  Before going ahead,  we    note  that the  heat engine is constructed into a closed path in the $P-V$ plane. It is recalled that  the heat absorbed is defined as $Q_H$ and the heat discharged is  given by $Q_C$ as represented in Fig.(\ref{cycleF}).
To  go  beyond  the previous   thermodynamical proprieties, we  compute  the specific heat at   constant volume and at  constant pressure. In particular, one  has 
\begin{equation}
C_{V}= T\left(\frac{\partial S}{\partial T}\right)_{V}= 0.
\label{45}
\end{equation}
However, the  heat capacity at constant pressure is  found to be
\begin{equation}
C_{p}= T\left(\frac{\partial S}{\partial T}\right)_{p}.
\end{equation}
This quantity is needed to  investigate   the  associated work from the heat energy according to a cycle between two sources (cold/hot) with  the temperatures $T_{cold}$ and $T_{hot}$, respectively.
Then, we make contact with the Carnot cycle defined as a simple cycle described by two isobars and two isochores as reported  in \cite{30}. These  configurations are illustrated in Fig.(\ref{fz}). In particular,  we  examine the  DE  effect on the efficiency of the  Hayward-AdS black holes, as  represented  in Figure.(\ref{cycleF}), The heat engine is drown  by  a rectangular cycle $(1\longrightarrow 2 \longrightarrow 3 \longrightarrow 4 \longrightarrow 1)$ in the $P-V$ plane.\\

\begin{figure}[!ht]
		\centering
		\hspace{-0.5cm}
		 \includegraphics[scale=.5]{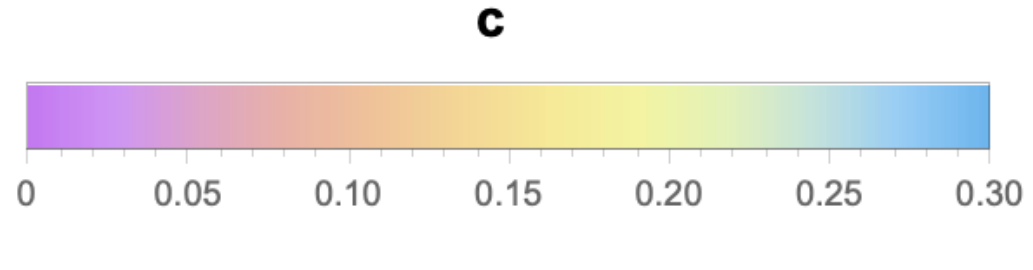}
			\begin{tabbing}
			\hspace{5.2cm}\= \hspace{5.2cm}\=\kill
			\includegraphics[scale=.41]{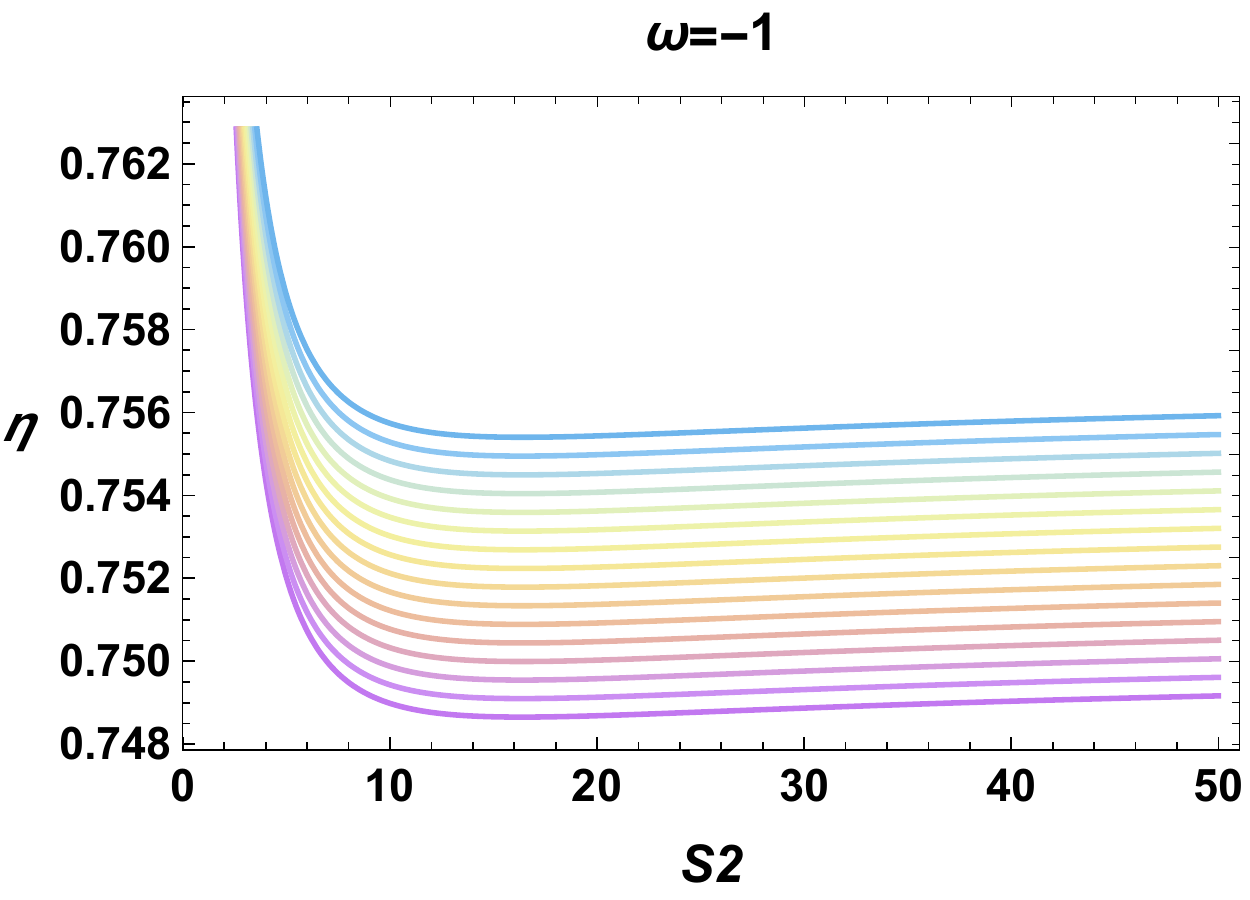} \>
			\includegraphics[scale=.41]{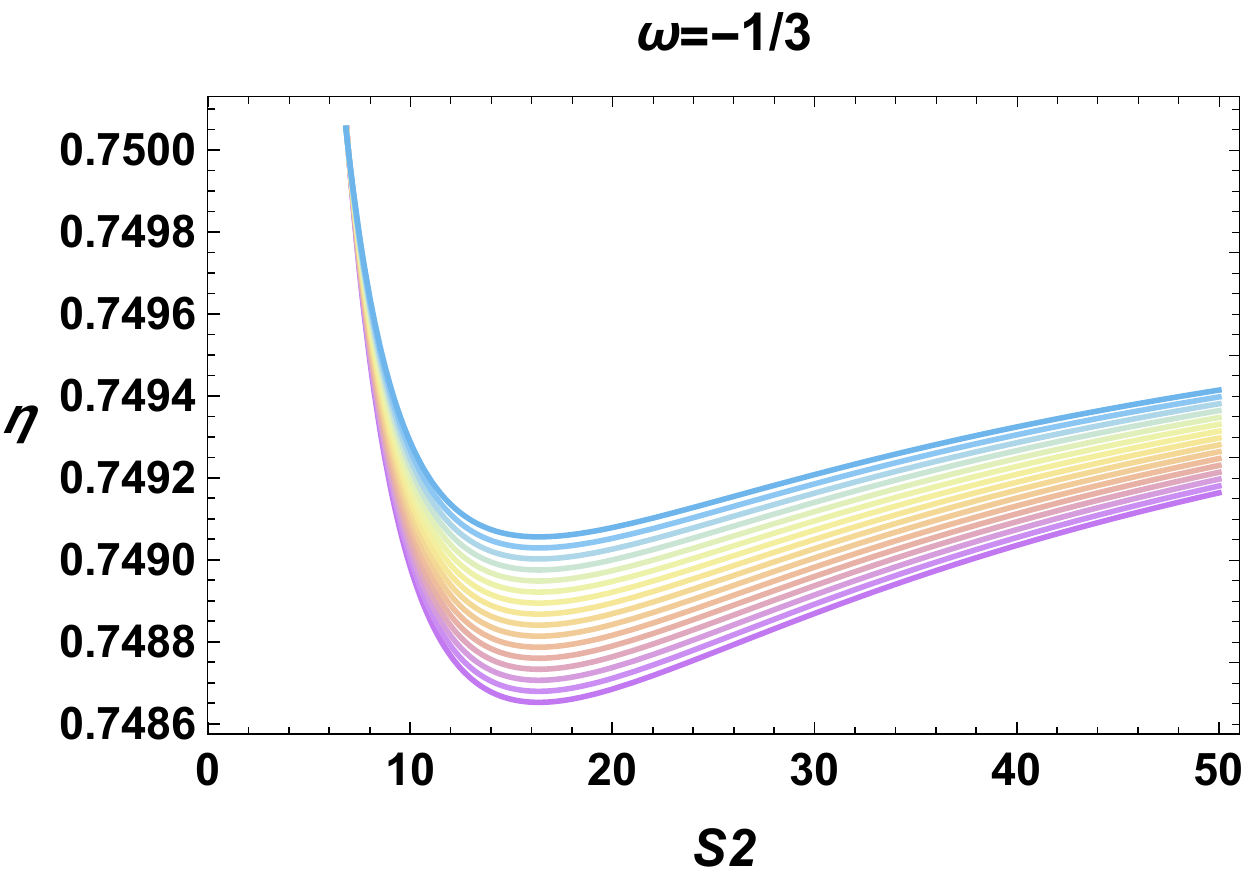} \>
			\includegraphics[scale=.41]{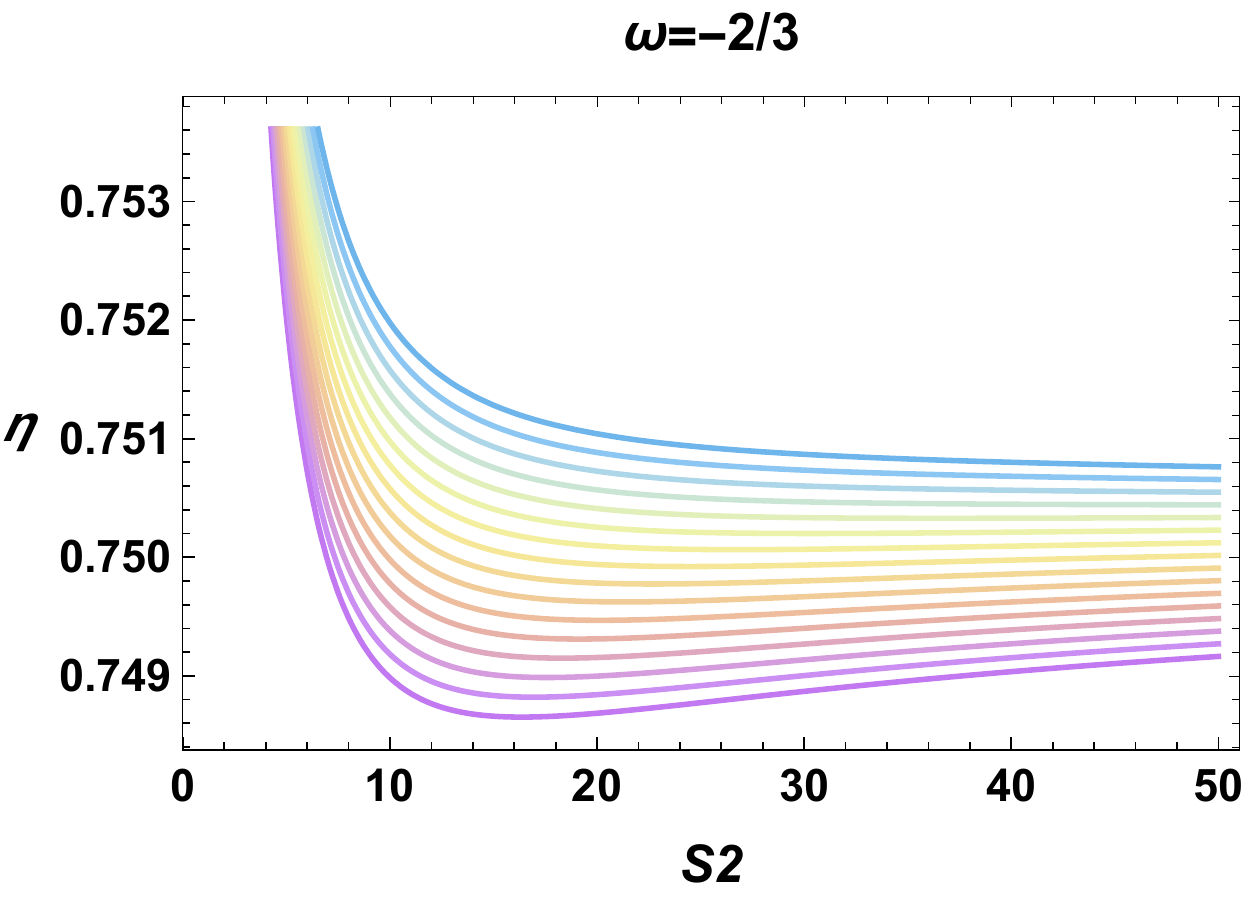} \\
			\includegraphics[scale=.41]{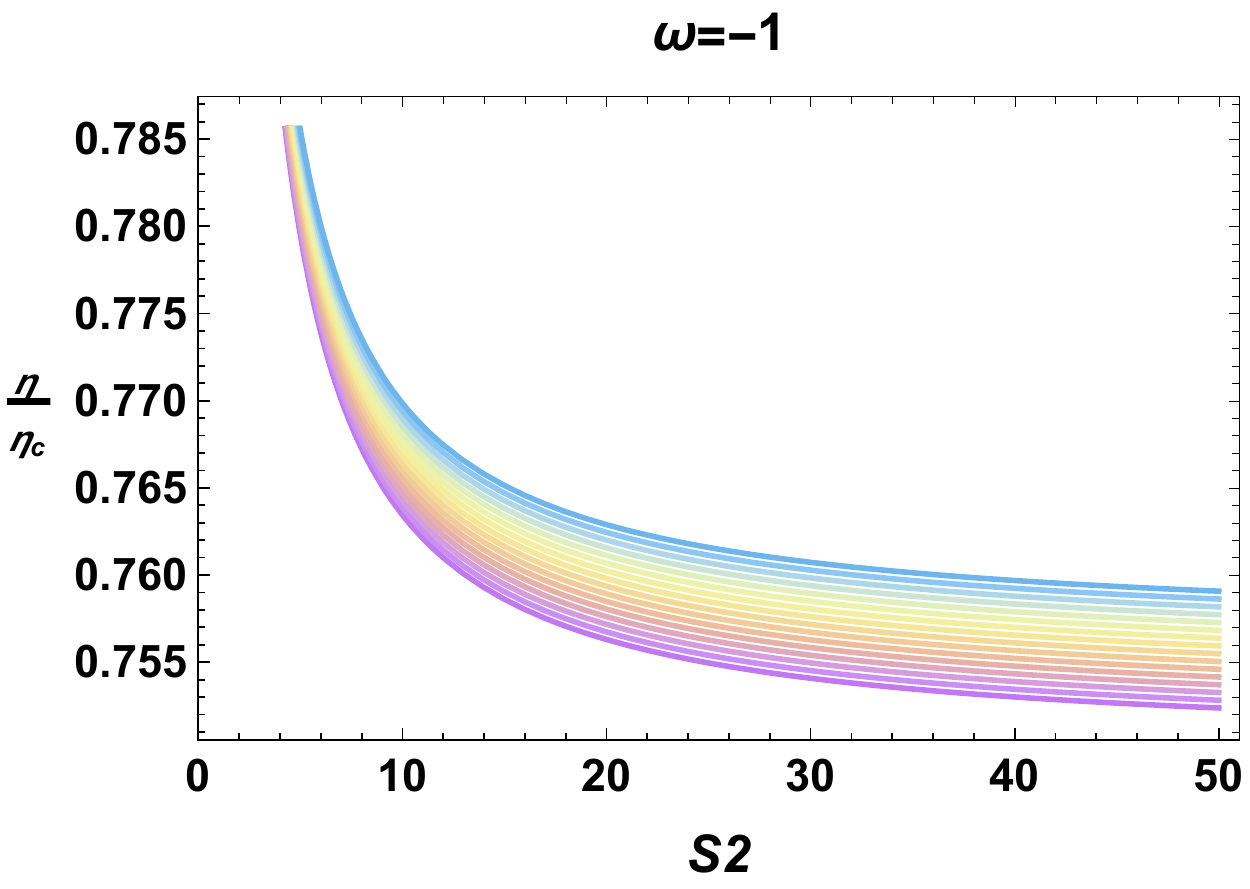} \>
			\includegraphics[scale=.41]{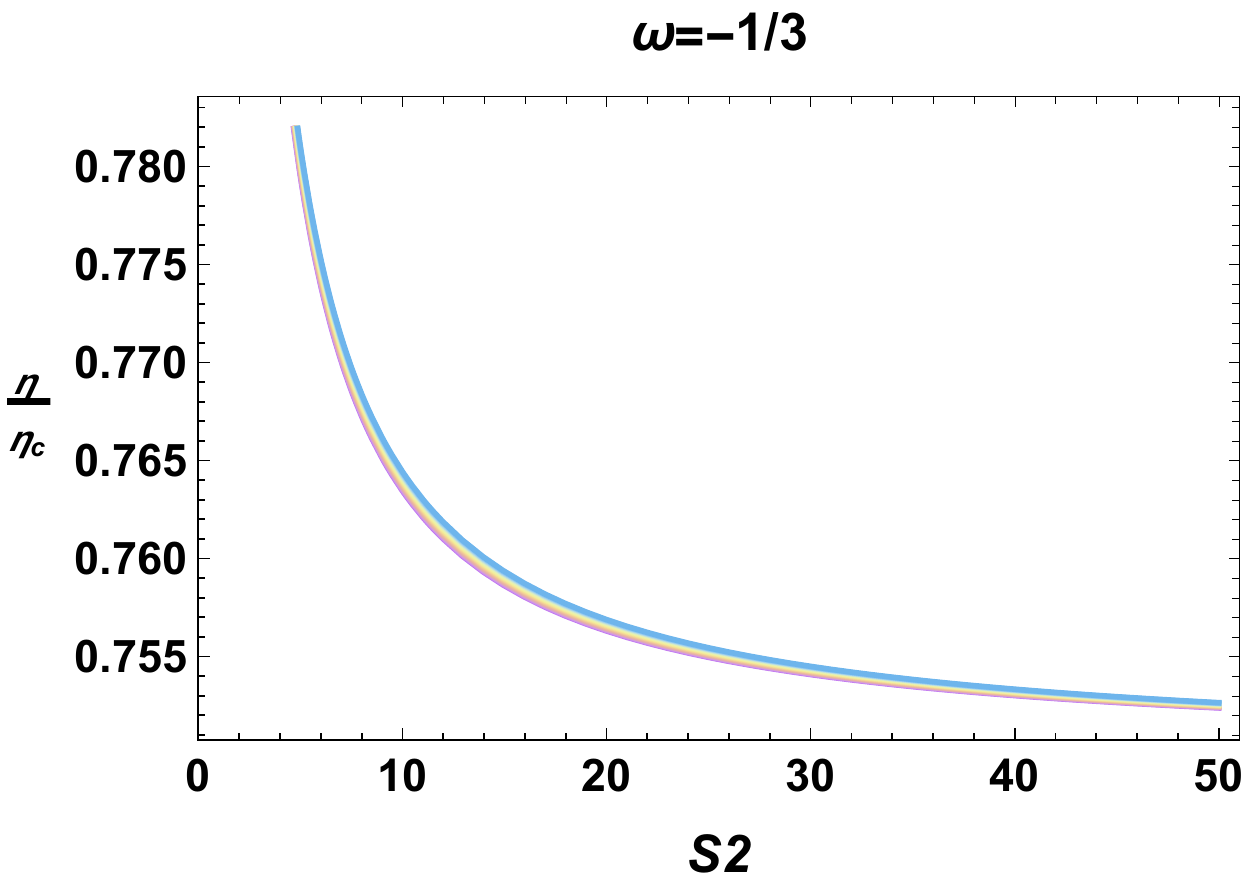} \>
			\includegraphics[scale=.41]{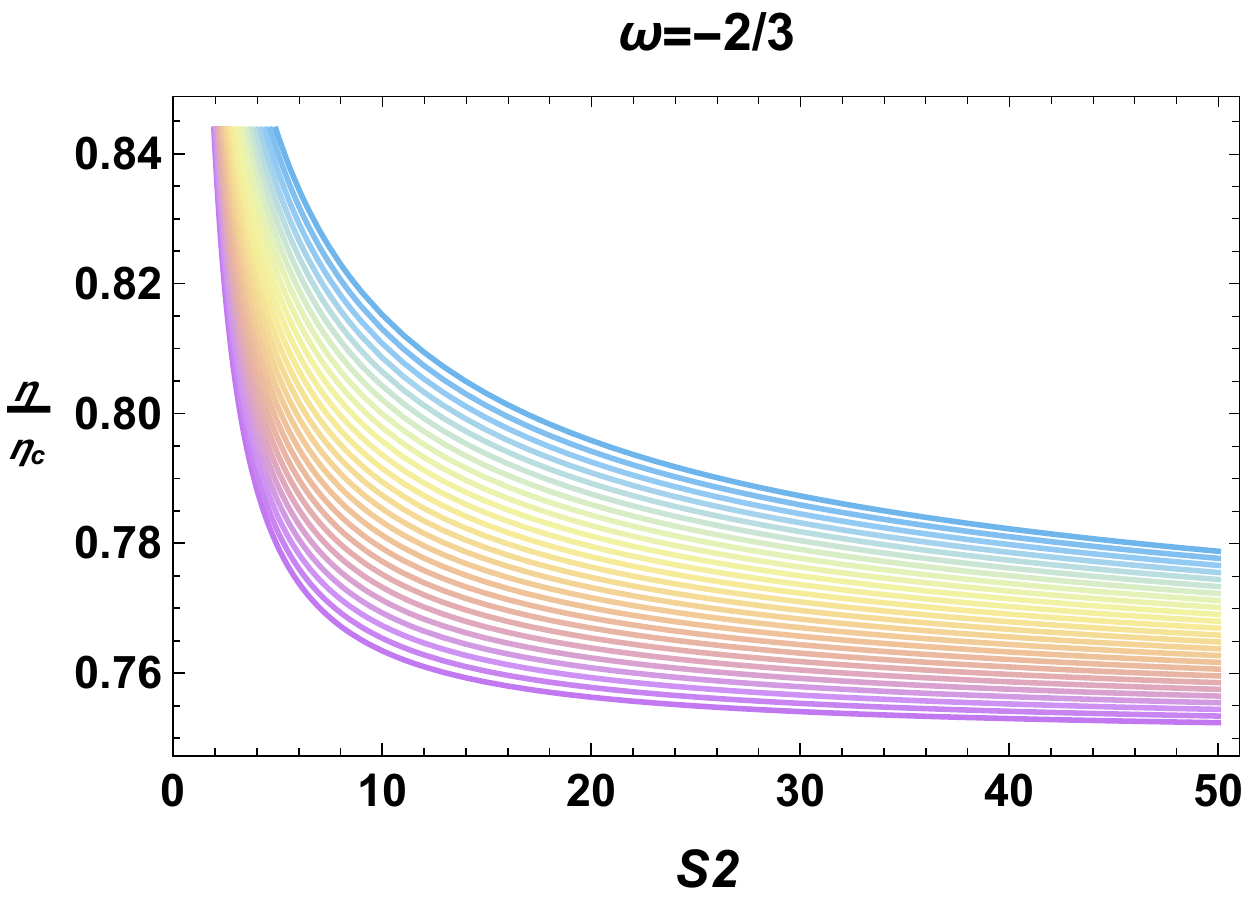} \\
			 \end{tabbing}
\vspace{-1cm}	
\caption{{\it \footnotesize  The functional relationship between the thermal engine efficiency $\eta$, the ratio $\eta/\eta_{car}$ and   the entropy $S_2$ of the Hayward-AdS surrounded by quintessence field for different values of $\omega_q$ and $c$. It has been taken  $Q=0.3$, $P_1=4$, $S_1=1$ and $P_4=1$.}}
\label{f2}
\end{figure}

\begin{figure}[!ht]
		\centering
		\hspace{-0.5cm}
		 \includegraphics[scale=.4]{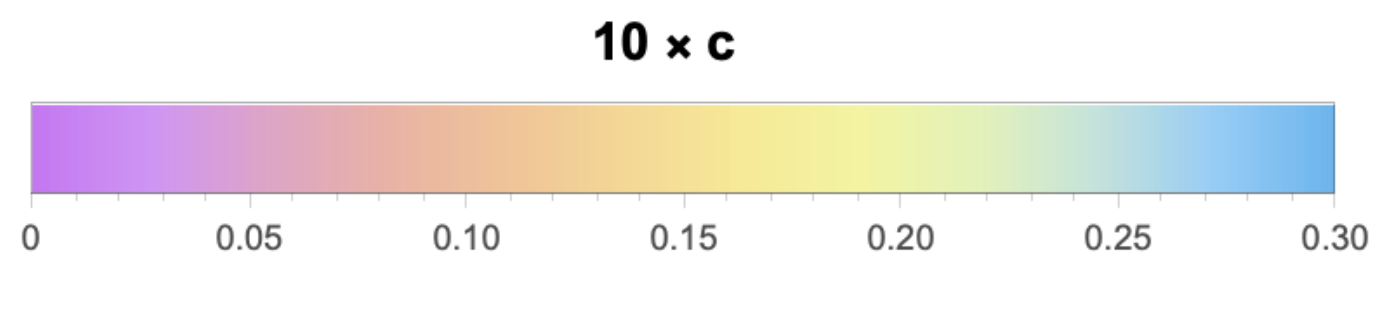}
			\begin{tabbing}
			\hspace{5.2cm}\= \hspace{5.2cm}\=\kill
			\includegraphics[scale=.41]{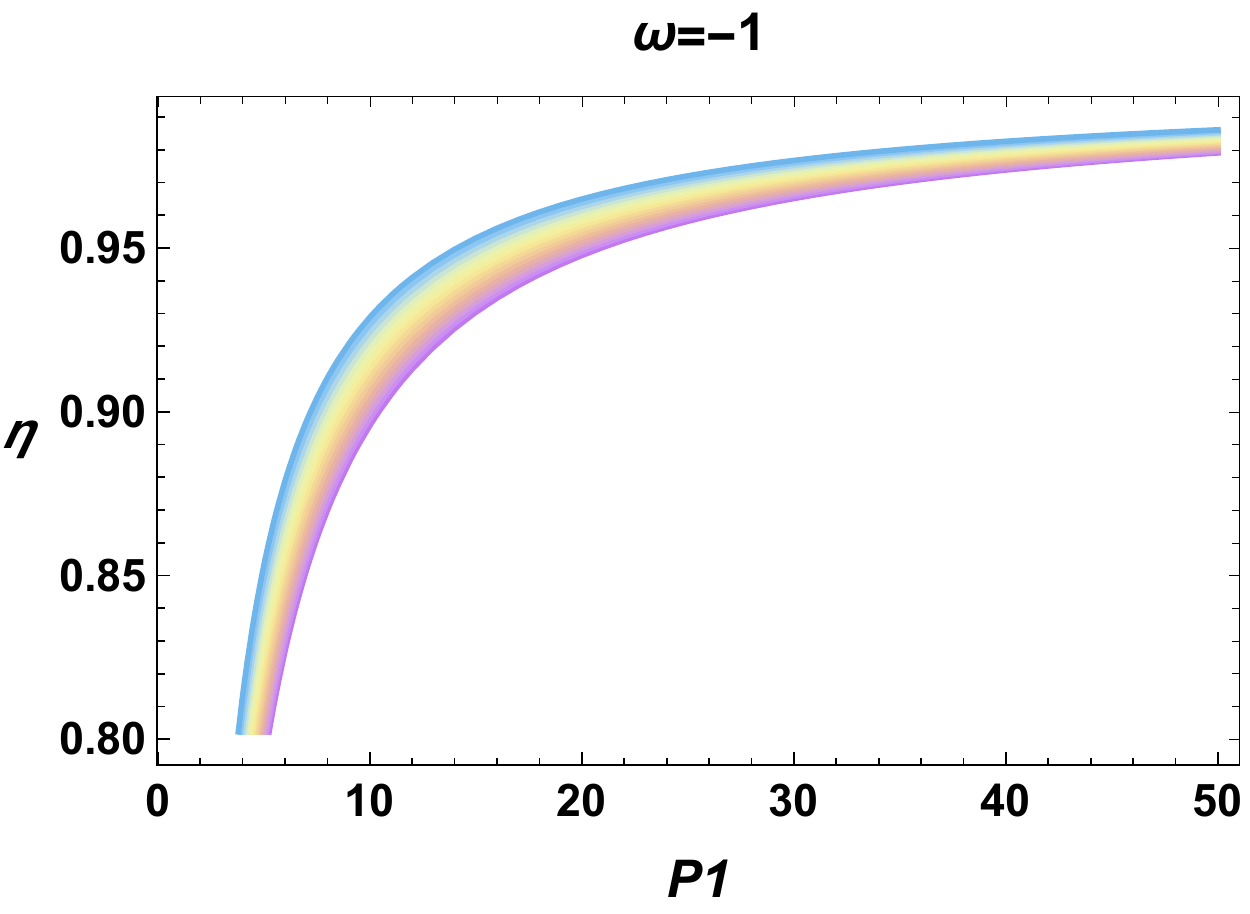} \>
			\includegraphics[scale=.41]{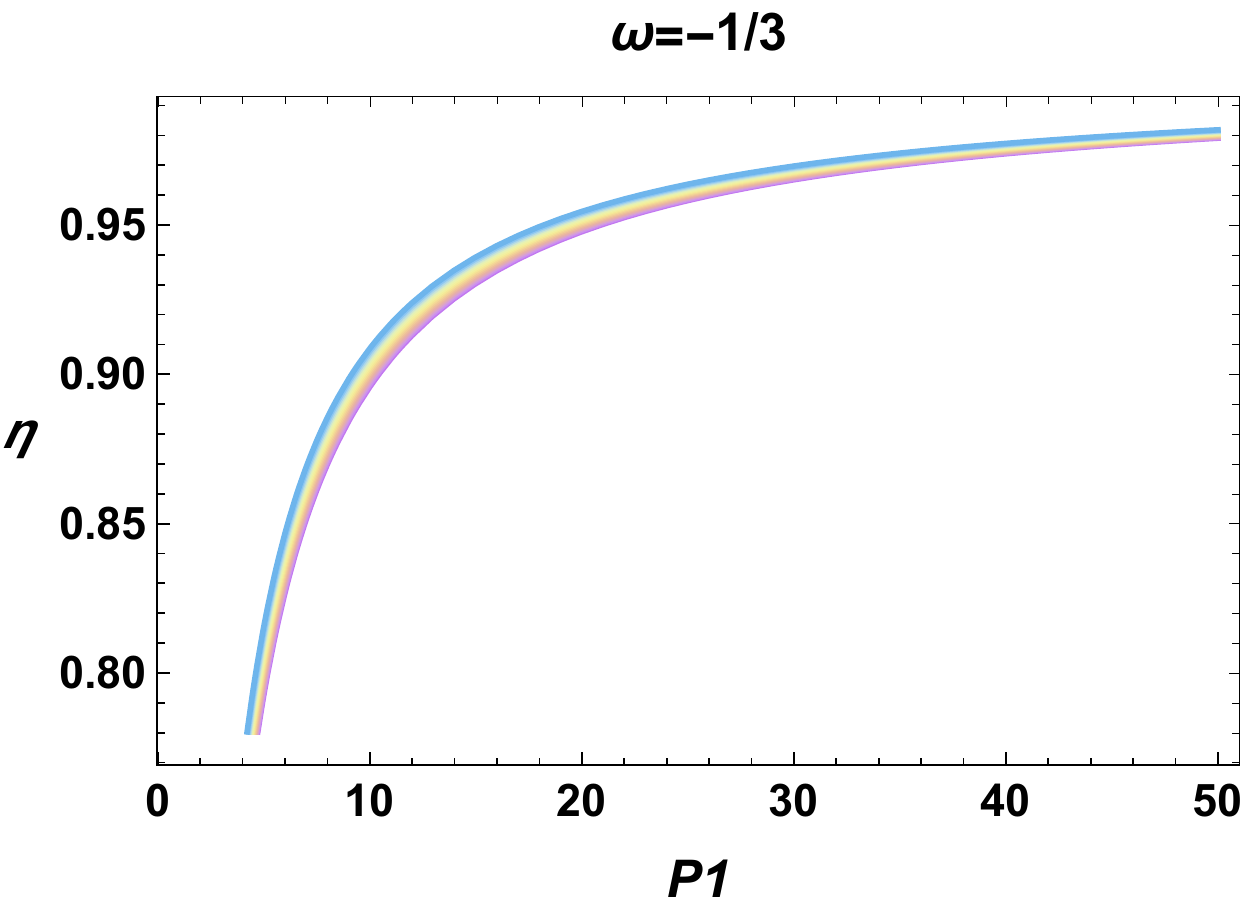} \>
			\includegraphics[scale=.41]{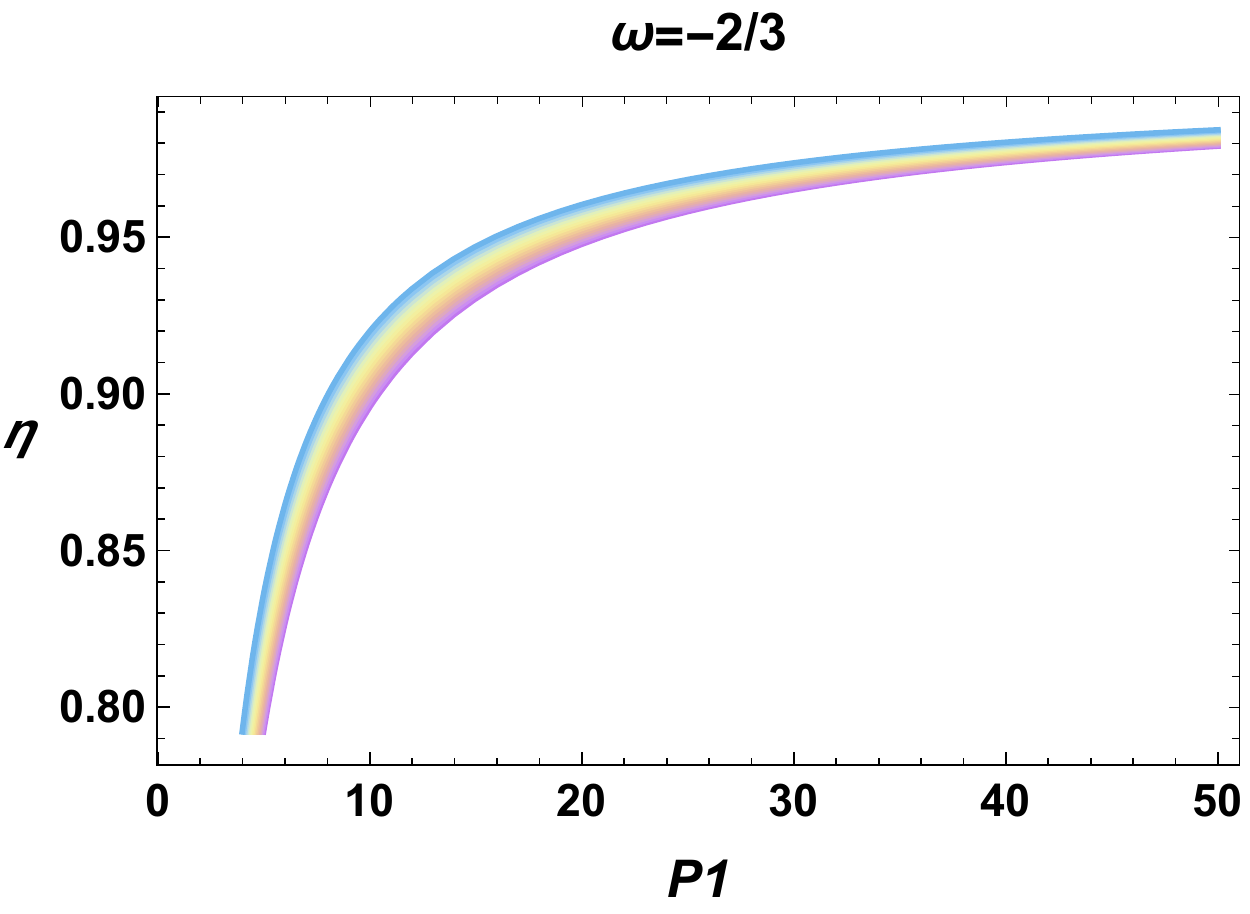} \\
			\includegraphics[scale=.41]{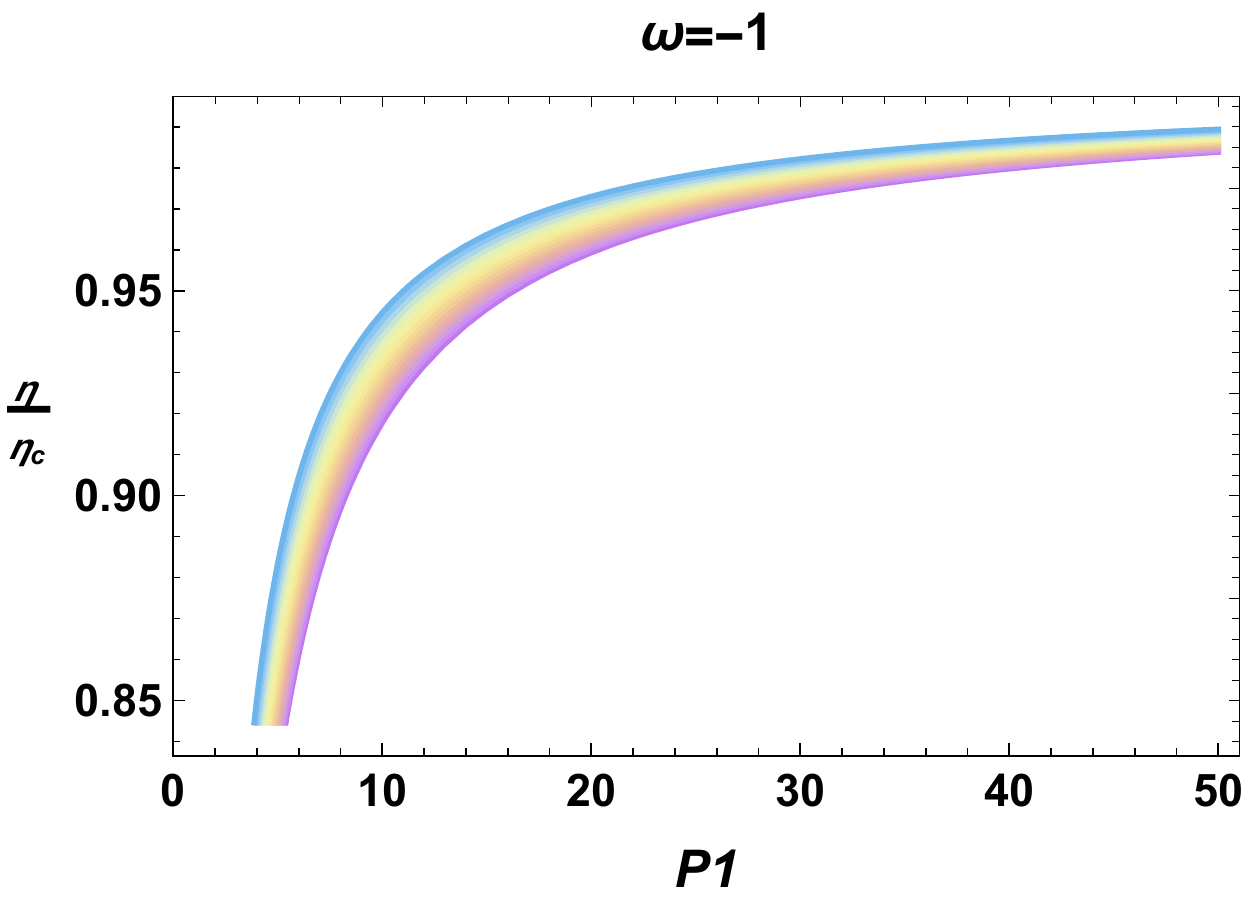} \>
			\includegraphics[scale=.41]{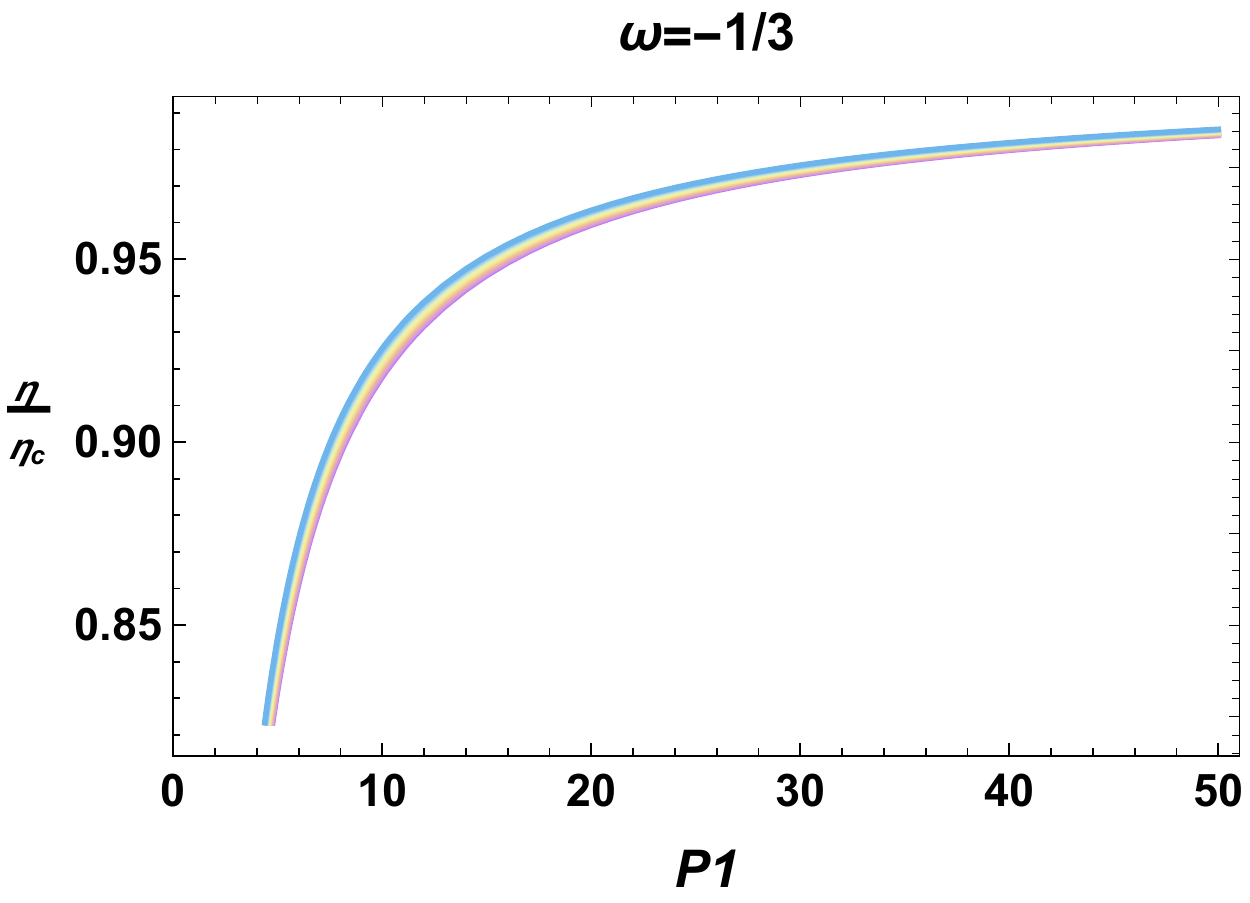} \>
			\includegraphics[scale=.41]{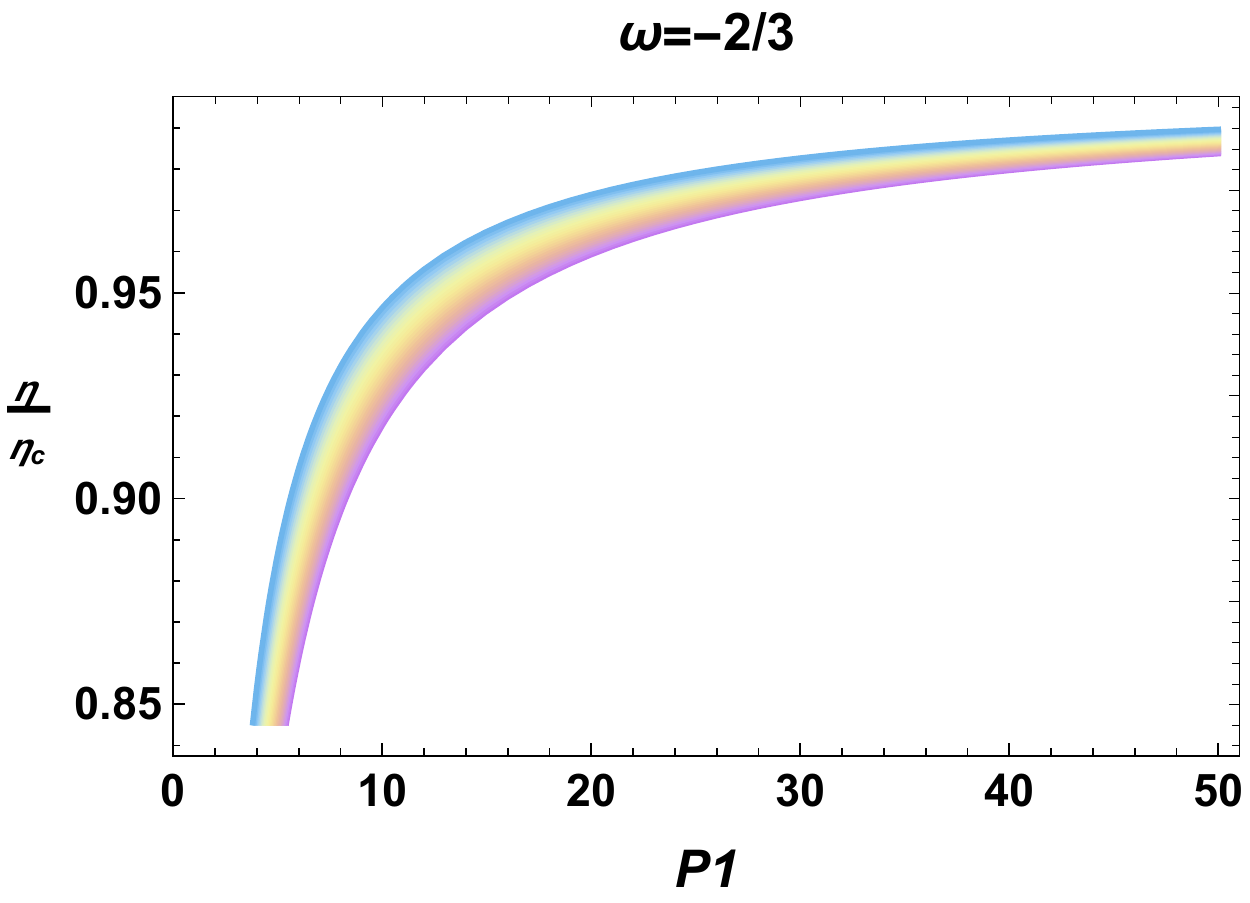} \\
		          \end{tabbing}
		          \vspace{-1cm}	
\caption{{\it \footnotesize  The functional relationship between the thermal engine efficiency $\eta$, the ratio $\eta/\eta_{car}$ and the pressure $P_1$ of the Hayward-AdS surrounded by quintessence field for different values of $\omega_q$ and $c$.  It has been  taken $Q=0.1$, $S_1=1$, $P_4=1$ and $S_2=4$. }}
\label{f221}
\end{figure}

The output work $W$, being the area of  the rectangle, is given by
\begin{align}
W=\oint PdV = P_{1}\left(V_{2}-V_{1}\right)+P_{4}\left(V_{4}-V_{3}\right).
\end{align}
Developing the calculations, one obtains
\begin{align}
W=\frac{4}{3\sqrt{\pi}}\left({S_{2}}^{\frac{3}{2}}-{S_{1}}^{\frac{3}{2}}\right)\left(P_{1}-P_{4}\right).
\end{align}
Since  $C_V=0$, we should compute the heat $Q_H$ during the process $1\longrightarrow2$. Indeed, the heat $Q_H$  can be  calculated using the following integral
\begin{equation}
Q_{H}= \int_{T_{1}}^{T_{2}}C_{p}\left(P_{1},T\right)dT = \int_{S_{1}}^{S_{2}}\left(\frac{\partial T}{\partial S}\right) dS = \int_{S_{1}}^{S_{2}} T dS = M_{2}-M_{1}.
\end{equation}
Using  Eq.(\ref{1}),  we obtain
\begin{equation}
Q_{H}= \frac{6 c S \pi ^{\frac{3 \omega_q }{2}} \left(\pi ^2 Q^3 (\omega_q +1)+\sqrt{\pi } S^{\frac{3}{2}} \omega_q \right)+2 S^{\frac{3 (\omega_q +1)}{2}} \left(S^{\frac{3}{2}} (8 P S+1)-2 \pi ^{\frac{3}{2}} Q^3\right)}{S^{\frac{3 \omega_q }{2}} \left(S^2 (8 P S-1)+8 \pi ^{\frac{3}{2}} Q^3 \sqrt{S}\right)-3 c \pi ^{\frac{3 \omega_q }{2}+\frac{1}{2}} \left(\pi ^{\frac{3}{2}} Q^3 \left(3 \omega_q ^2+8 \omega_q +5\right)+S^{\frac{3}{2}} \omega_q  (3 \omega_q +2)\right)}.
\end{equation}
Dividing the work  by this expression,  we can   get  the efficiency. Indeed, it is given by
\begin{equation}
\eta=\frac{W}{Q_{H}} = \frac{8 ({P_1}-P_4) \left({S_2}^{\frac{3}{2}}-{S_1}^{\frac{3}{2}}\right)}{\beta_{S1}  {S_1}^{-\frac{3}{2}  (\omega_q +1)} \left(3 c \pi ^{\frac{3 \omega_q }{2}+\frac{1}{2}}-\alpha_{S1} \right)+\beta_{S2}  {S_2}^{-\frac{3}{2}  (\omega_q +1)} \left(\alpha_{S2} -3 c \pi ^{\frac{3 \omega_q }{2}+\frac{1}{2}}\right)}
\label{456}
\end{equation}
where the entropy functions $\alpha_S$ and $\beta_S$ take the following form
\begin{eqnarray}
\alpha_{S} & = & (8 P_1 S+3) S^{\frac{3 \omega_q }{2}+\frac{1}{2}}, \\
\beta_{S} & = & \pi ^{\frac{3}{2}} Q^3+S^{\frac{3}{2}}.
\end{eqnarray}
 Considering $Q=0$,  we  recover  a similar equation of $\eta$  as reported in \cite{30}. A close  examination shows that small c expansions provide the following expression
\begin{equation}
\label{ }
\eta\sim\eta(c=0)+\eta(c)+O\left(c^2\right),
\end{equation}
where $\eta(c=0)$ and $\eta(c)$ are giving by
\begin{eqnarray}
\eta(c=0) & = &\frac{8 {S_1} {S_2} ({P_1}-{P_4}) \left({S_2}^{\frac{3}{2}}-{S_1}^{\frac{3}{2}}\right)}{\beta_{S2}  (8 {P_1} {S_2}+3){S_1}-\beta_{S1} (8 {P_1} {S_1}+3){S_2}}, \\
\eta(c) & = &\frac{8 c \left((\gamma_{S1}-\gamma_{S2}) ({P_1}-{P_4}) \left({S_2}^{\frac{3}{2}}-{S_1}^{\frac{3}{2}}\right)\right)}{\left(\alpha_{S2}\beta_{S2}{S_2}^{-\frac{3}{2}
(\omega_q+1)}-\alpha_{S1}\beta_{S1}{S_1}^{-\frac{3}{2}(\omega_q+1)}\right)^2}
\end{eqnarray}
and where one has  used  $\gamma_{S}=3 \beta_S  \pi ^{\frac{3 \omega_q }{2}+\frac{1}{2}} S^{\frac{1}{2} (-3) (\omega_q +1)}$. This expression should be  compared  with the Carnot efficiency $\eta_{car}$  by  investigating  the ratio $\frac{\eta}{\eta_{car}}$.  Indeed, the  efficiency of the Carnot engine is given by 
\begin{equation}
\eta_{car}=1-\frac{T_{cold}}{T_{hot}} = 1 - \frac{T\left(P_{4},S_{1}\right)}{T\left(P_{1},S_{2}\right)}.
\end{equation}
By employing the previous equations, the efficiency $\eta$ can take the following form
\begin{equation}
\eta_{car}= 1-\frac{ S_2^{\frac{1}{2}(3 \omega_q +1)}\beta_{S2} (\lambda_{S1P4}+c\, \delta_{S1})}{S_1^{\frac{1}{2}(3 \omega_q +1)}\beta_{S1} (\lambda_{S2P1}+c\, \delta_{S2})},
\label{789}
\end{equation}
where one has
\begin{equation}
\lambda_{SP}=S^{\frac{1}{2}(3 \omega_q+1)}  \left(S^{\frac{3}{2}} (8 PS+1)-2 \pi ^{\frac{3}{2}} Q^3\right)
\end{equation}
and where the entropy function $\delta_{S}$ is given by
\begin{equation}
\delta_{S}  =  3  \pi ^{\frac{3 \omega_q }{2}}  \left(\pi ^2 Q^3 (\omega_q +1)+\sqrt{\pi } S^{\frac{3}{2}} \omega_q \right).
\end{equation}
Carrying out small c expansions, the Carnot  efficiency $\eta_{car}$  reduces to
\begin{equation}
\label{ }
\eta_{car}\sim\eta_{car}(c=0)+\eta_{car}(c)+O\left(c^2\right),
\end{equation}
where the involved terms are
\begin{eqnarray}
\eta_{car}(c=0) & =&1-\frac{ S_2^{\frac{1}{2}(3 \omega_q +1)}\beta_{S2} (\lambda_{S1P4}+c\, \delta_{S1})}{S_1^{\frac{1}{2}(3 \omega_q +1)}\beta_{S1} (\lambda_{S2P1}+c\, \delta_{S2})},\\
\eta_{car}(c) & = &-\frac{ S_2^{\frac{1}{2}(3 \omega_q +1)}\beta_{S2} (\delta_{S1}\lambda_{S2P1}-\delta_{S2}\lambda_{S1P4})c}{S_1^{\frac{1}{2}(3 \omega_q +1)}\beta_{S1} (\lambda_{S2P1})^2}.
\end{eqnarray}
It follows from these calculations that the heat engine efficiency depends on the involved quantities including the  dark energy density, the state  parameter, the entropy and the pressure. To visualize  clearly the associated behaviors, we consider   the variations of  the  efficiency in terms of various quantities. In particular, we consider three  models corresponding to  the known values of the state  parameter  $\omega=-1, -\frac{2}{3}, -\frac{1}{3}$.\\
Concretely, Fig.(\ref{f2}) illustrates the variations of the efficiency $\eta$ and the ratio $\frac{\eta}{\eta_{car}}$   with respect to $S_2$ by taking different values of the  state parameter.
First, we analyze  the variation of  the efficiency $\eta$.     For $\omega_q=-1$ and $\omega_q=-\frac{2}{3}$, the efficiency decreases by increasing the entropy $S_2$ and  by decreasing $c$. Moreover, it  is almost constant for large  entropy $S_2$ for different values of $c$. For $\omega=-\frac{1}{3}$,  however, the behavior is quite different. The efficiency of the heat engine decreases for a certain range of the entropy. After reaching specific  values, it increases  with  $S_2$. Varying $S_2$, the efficiency increases  by increasing  the  DE filed intensity $c$.  However, it is  observed that the variation $\eta$ as a  function of $S_{2}$ presents a minimum, which increases  by increasing $c$.   However, this  minimum is not observed in the $\eta-S_2$ plane  for  the quintessential charged AdS black holes\cite{30}. The variation of $\frac{\eta}{\eta_{car}}$  in terms of  $S_2$ involves similar properties  for the three  $\omega$ models. In particular,  it  decreases by increasing $S_2$. For $\omega_q=-1$ and $\omega_q=-\frac{1}{3}$,  we observe that the field  intensity  $c$  provides  a non relevant effect on the ratio $\frac{\eta}{\eta_{car}}$. For  $\omega_q=-1$, however, the ratio $\frac{\eta}{\eta_{car}}$ increases by increasing  the DE filed intensity $c$.\\
Now, we move to examine the variation of the efficiency in terms of the pressure  $P_1$ by varying   $c$ and $\omega_q$. The associated behaviors  are  illustrated in Fig.(\ref{f221}) by scaling such a parameter as $10c$.

\begin{figure}[ht!]
		\begin{center}
		\centering
			\begin{tabbing}
			\centering
			\hspace{7.6cm}\=\kill
			\includegraphics[scale=.55]{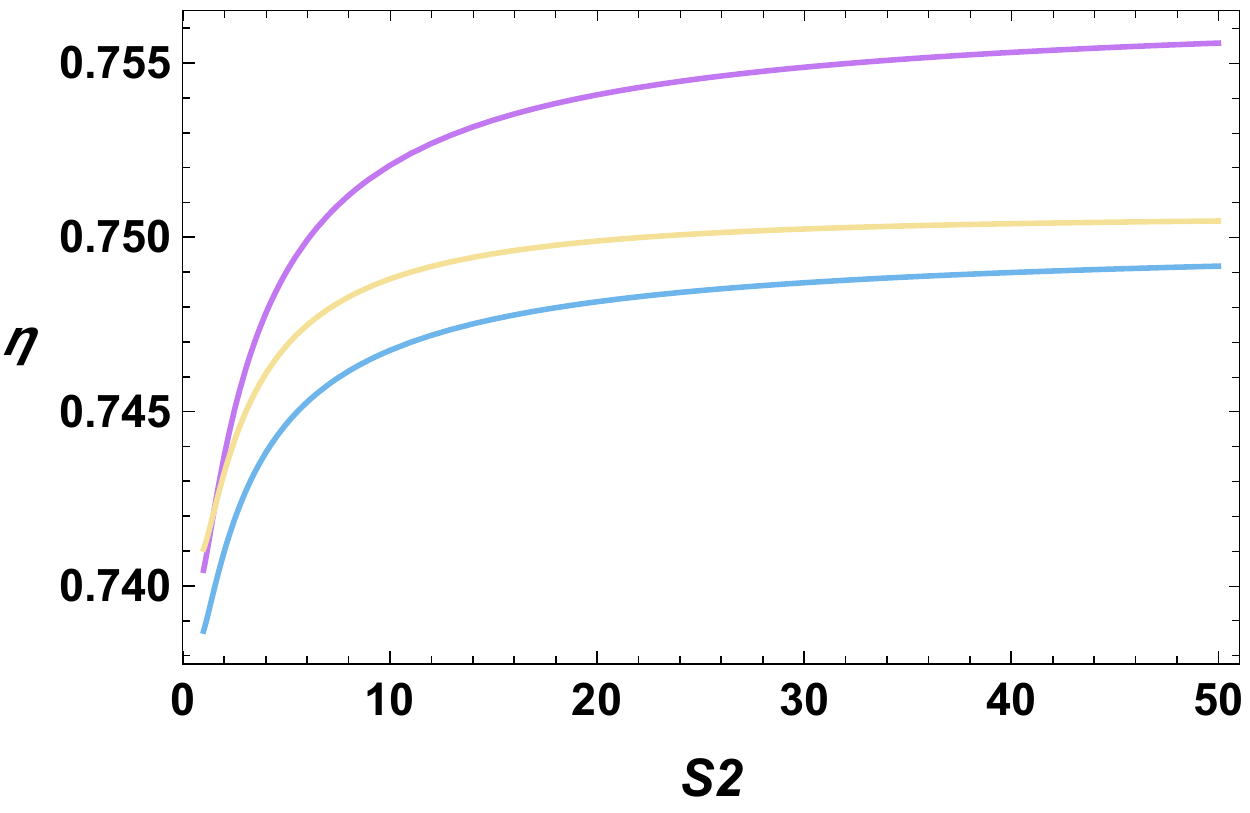} \>
			\includegraphics[scale=.55]{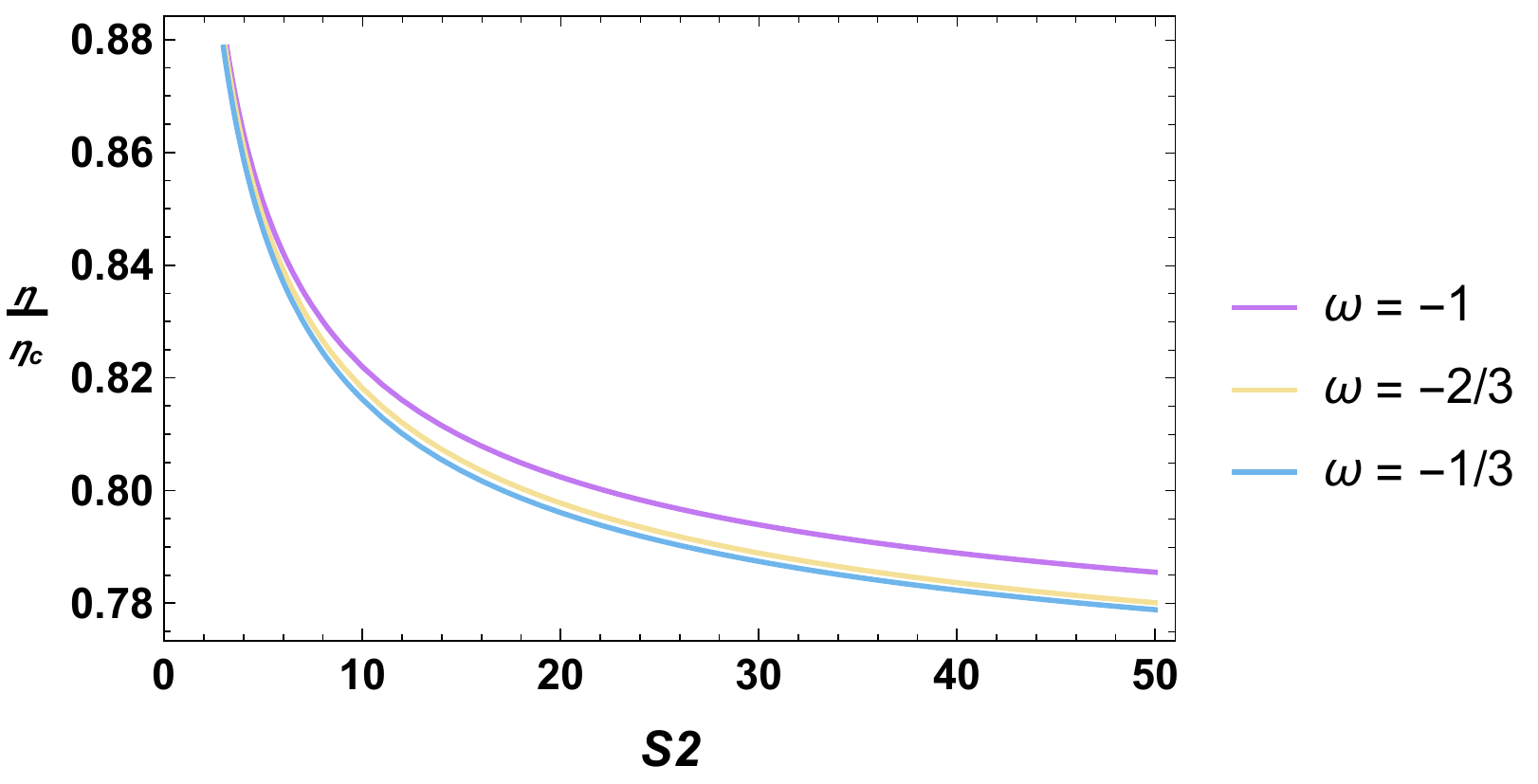} \\
		   \end{tabbing}
		   \vspace{-1cm}	
\caption{{{\it \footnotesize The variation of efficiency $\eta$ and the ratio $\eta/\eta_{car}$ as function of $S_2$ for different values of $\omega_q$ by taking $c=0.3$, $Q=0.3$, $P_1=4$, $S_1=1$ and $P_4=1$.}}}
\label{f1}
\end{center}
\end{figure}

It has been remarked that  $\eta$ and $\frac{\eta}{\eta_{car}}$  involve  similar aspects.  For the  $P_1$ variation,  all panels obviously reveal  that always there is a monotonous trend of an  increase. Moreover,  the efficiency of the  heat engine  approaches  the maximum possible value $1$ by augmenting the  pressure $P_1$. Taking $\omega_q=-\frac{1}{3}$,  the effect of   $c$ on  the efficiency $\eta$ and the ratio  $\frac{\eta}{\eta_{car}}$  is   negligible. For $\omega_q=-1$ and $\omega_q=-\frac{2}{3}$, however,    they  increase  by increasing $c$. \\
To see the effect of $\omega_q$ on such behaviors, we plot in  Fig.(\ref{f1}) the variation  of such quantities  with respect to the entropy $S_ 2$  for three different values of the DE state parameter.
From Fig.(\ref{f1}), one can observe that the black hole cycle is more efficient  when   the parameter $\omega_q$ decreases. For  the ratio between  the efficiency  and the Carnot efficiency ($\frac{\eta}{\eta_{car}}$),   it has been seen  that, for  small values of the entropy $S_2$,  all the efficiencies are the same.   When the entropy becomes important with respect the pressure, the $\frac{\eta}{\eta_{car}}$ curves separate.  In particular, the ratio  $\frac{\eta}{\eta_{car}}$ increases  when   the parameter $\omega_q$ decreases.

\section{Optical aspect  of quintessential  Hayward-AdS black holes}
In this section, we investigate  the optical properties  of  the charged  quintessential Hayward-AdS black holes.   Precisely, we approach  the shadow geometrical configurations  in terms of  the involved parameters.
\subsection{Shadow behaviors}
\begin{figure}[!ht]
		\centering
			\begin{tabbing}
			\hspace{5.4cm}\= \hspace{5.4cm}\=\kill
			\includegraphics[scale=.35]{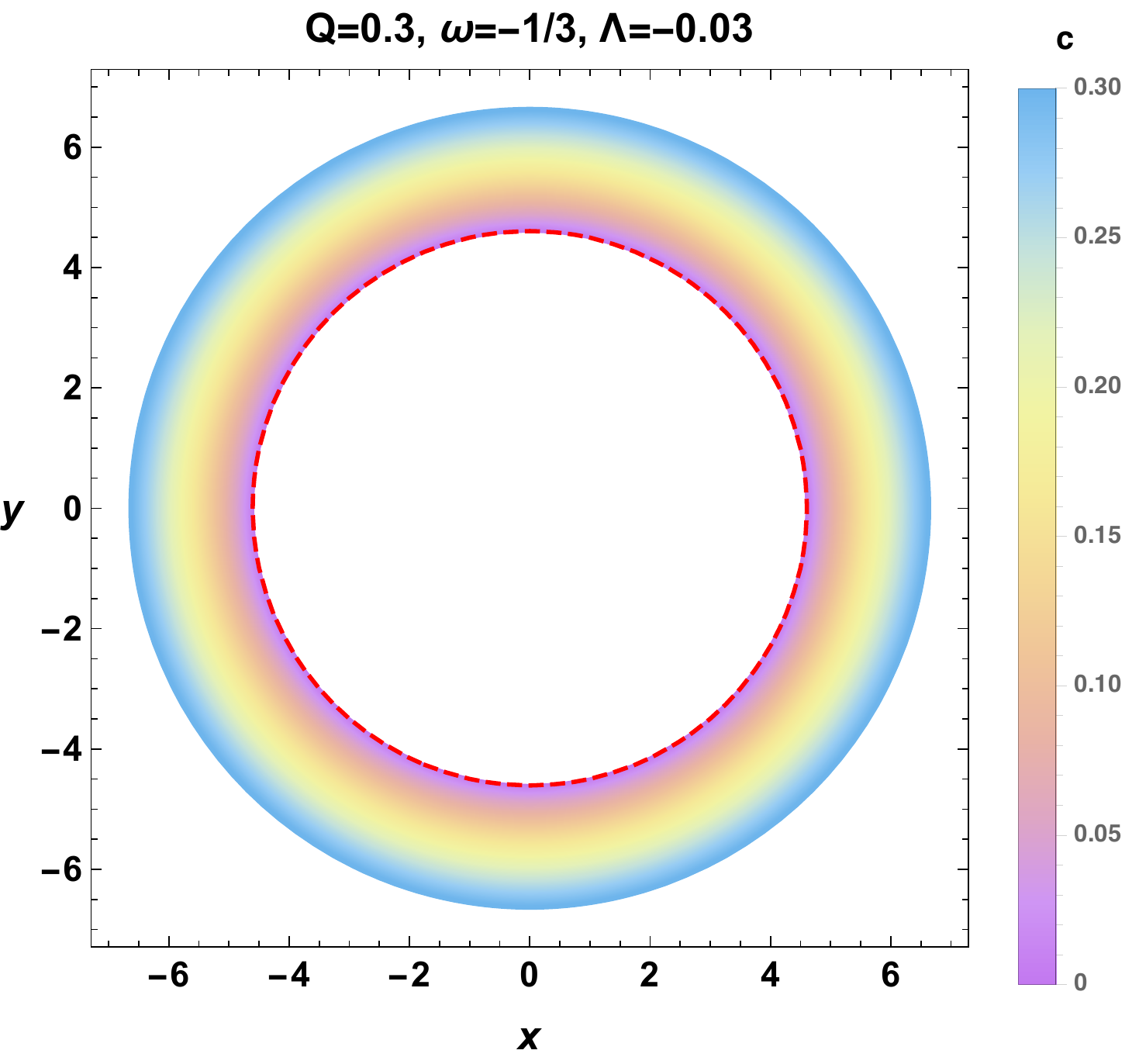} \>
			\includegraphics[scale=.35]{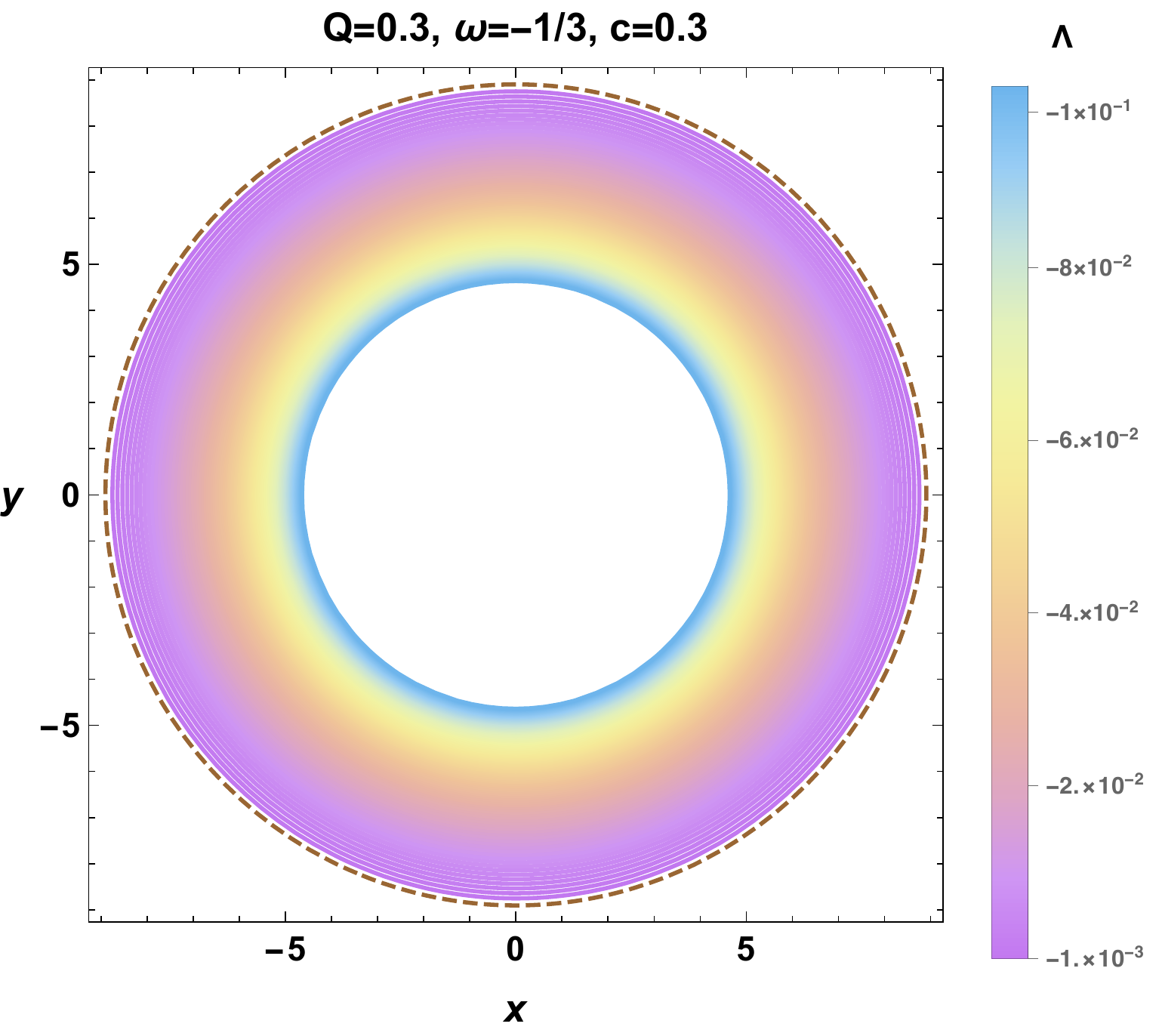} \>
			\includegraphics[scale=.35]{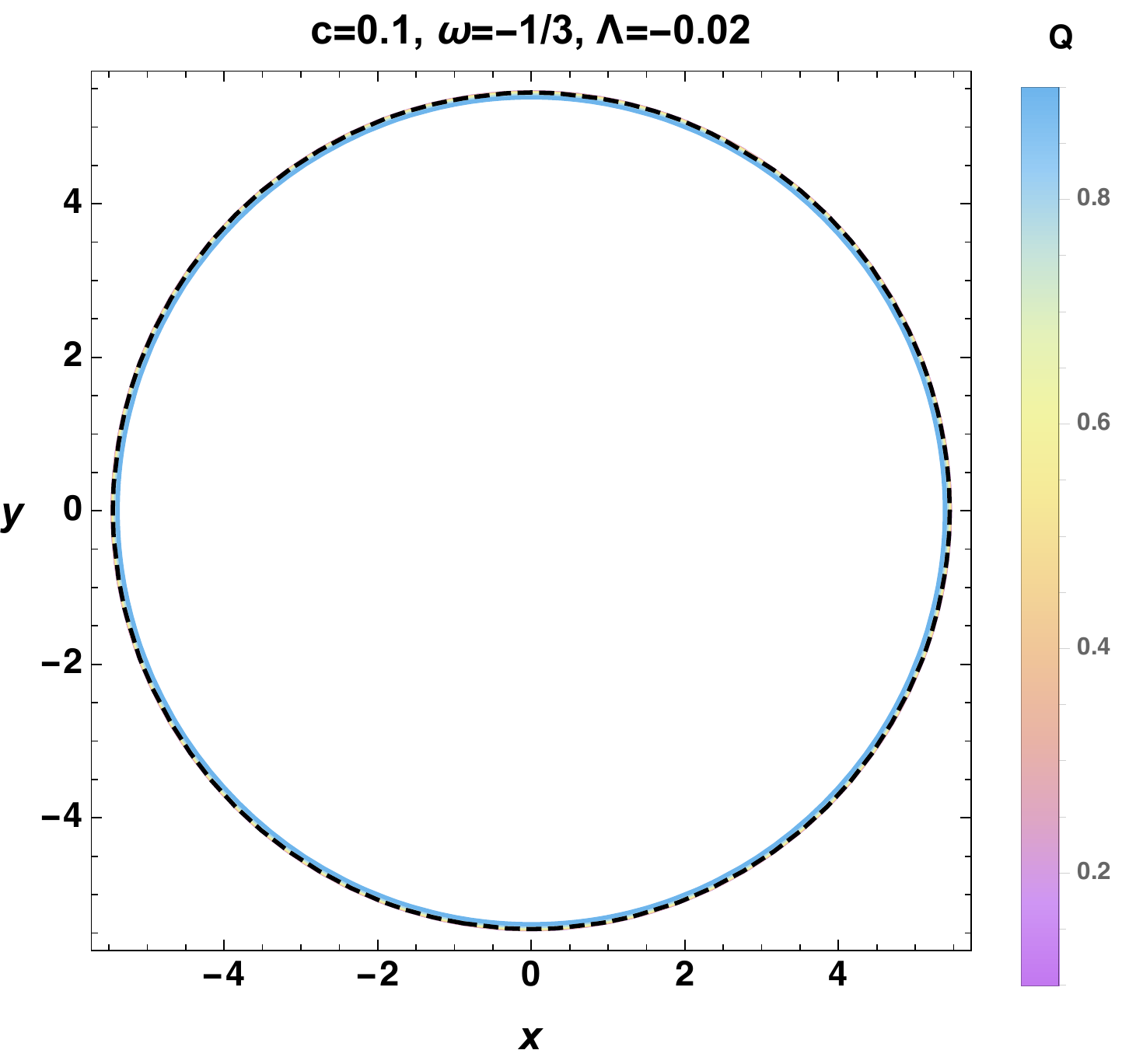} \\
		          \end{tabbing}
		          \vspace{-1cm}
\caption{\it \footnotesize  The  Hayward-AdS  black hole  shadows variation as function of $Q$, $\omega_q$ and $\Lambda$ in the celestial plane. Dashed and red,  brown and black circles correspond to $c=0$, $\Lambda=0$ and $Q=0$, respectively.}
\label{f222}
\end{figure}

\begin{figure}[!ht]
		\centering
			\begin{tabbing}
			\hspace{5.4cm}\= \hspace{5.4cm}\=\kill
			\includegraphics[scale=.39]{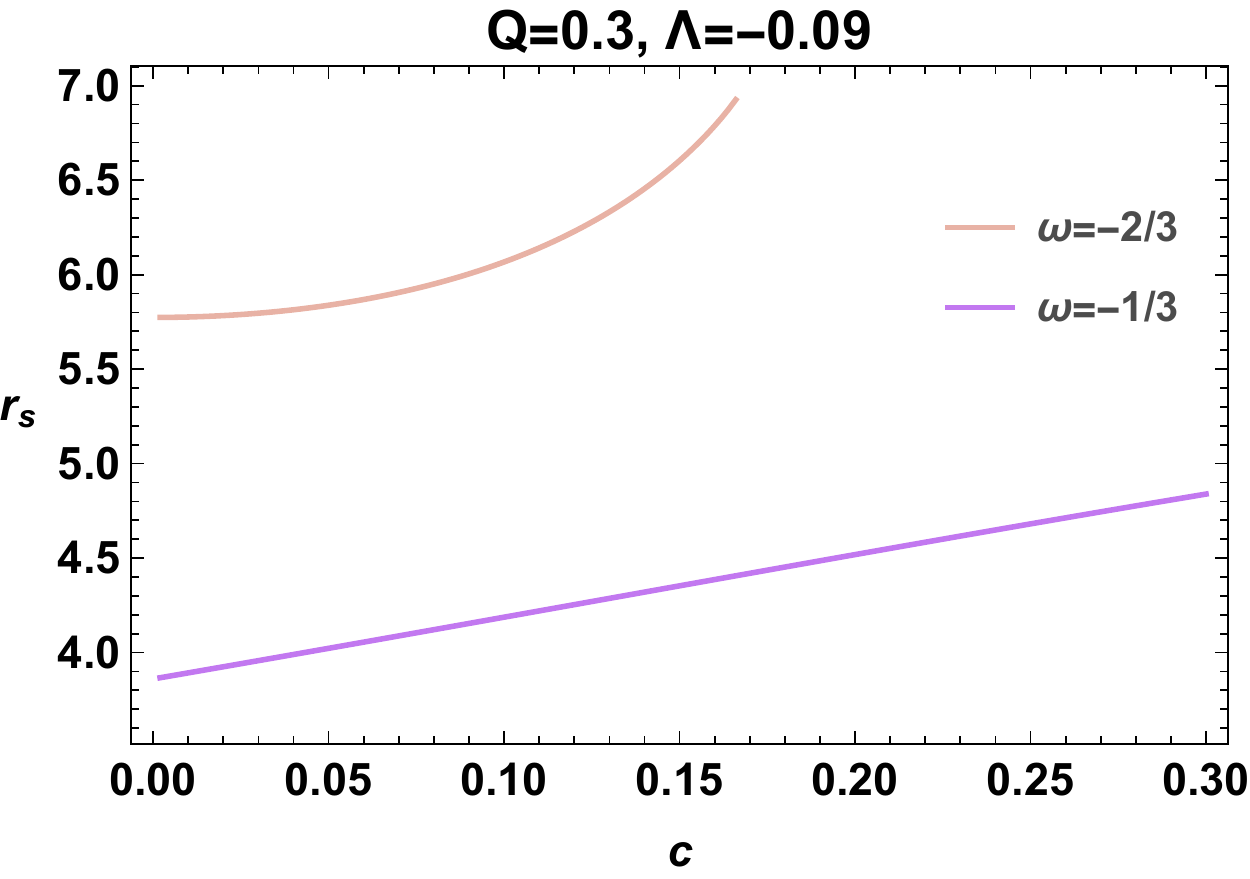} \>
			\includegraphics[scale=.39]{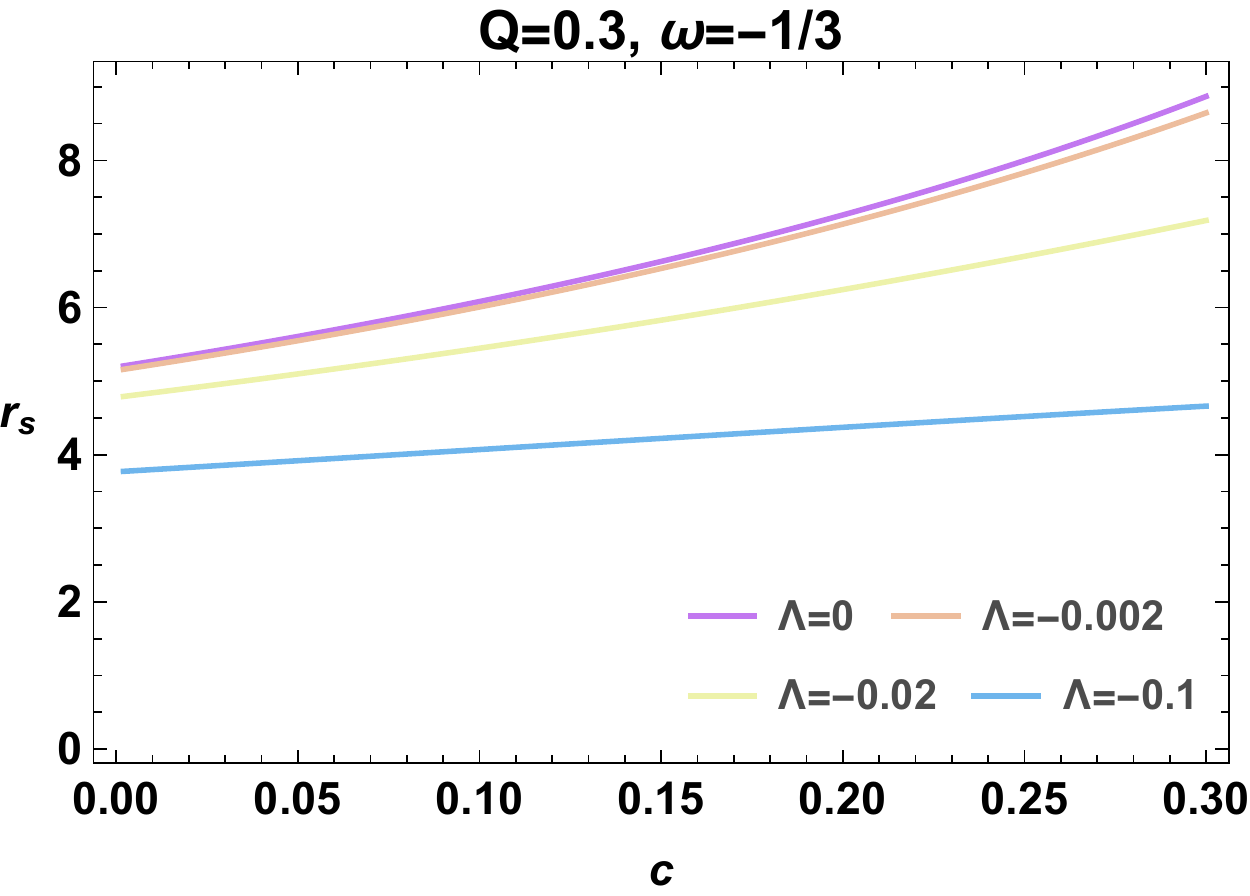} \>
			\includegraphics[scale=.39]{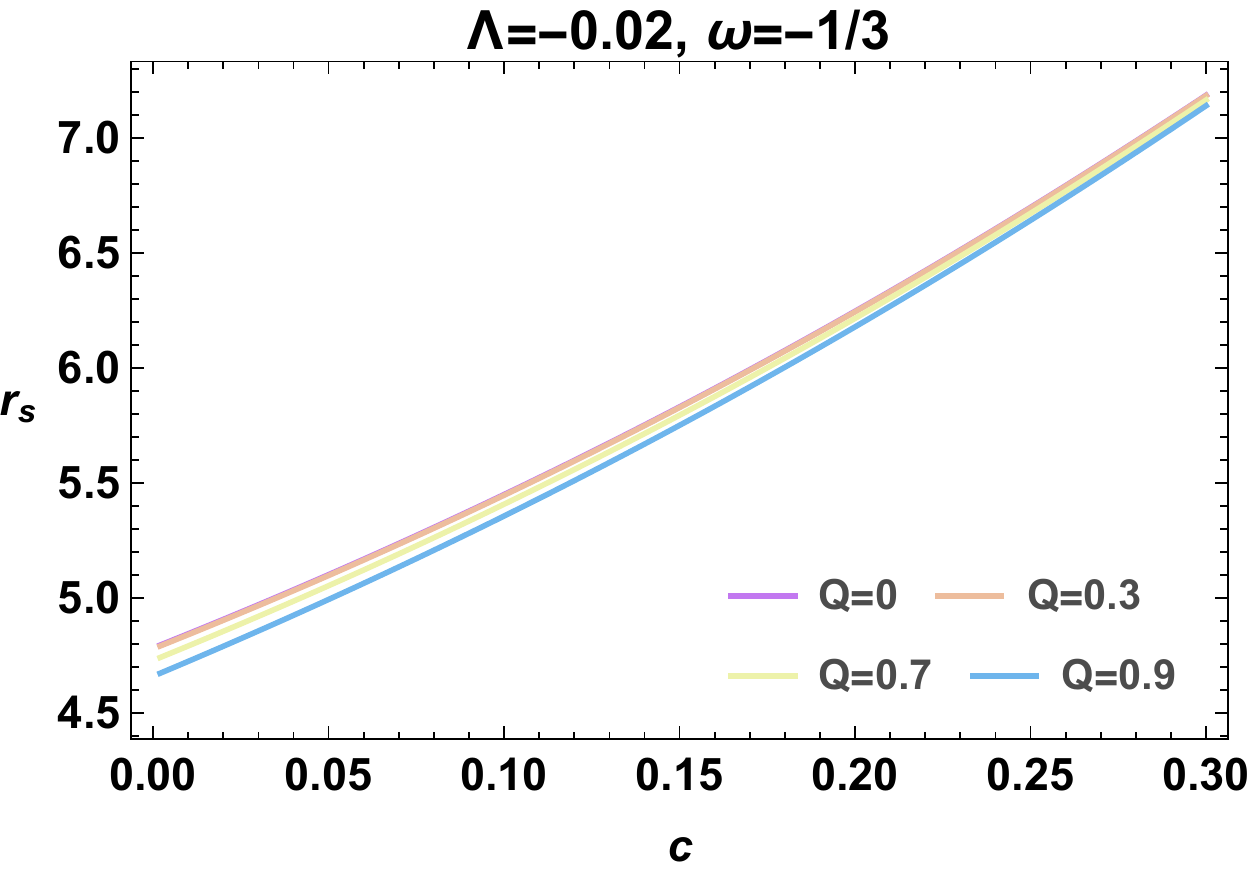} \\
		          \end{tabbing}
		           \vspace{-1cm}
\caption{\it \footnotesize The shadow radius variation  as function of $c$ parameter by varying $Q$, $\omega_q$ and $\Lambda$ parameters.  }
\label{F223}
\end{figure}

Following  to the activities  associated with  such  quintessential  AdS-black holes \cite{Carter:1968rr,Chandrasekhar}, the massless particle  equations of motion can be obtained  by  employing   the Hamilton-Jacobi method for a
photon in the associated  spacetime. The associated equation reads as 
\begin{equation}
\frac{\partial S}{\partial \tau }=-\frac{1}{2}g^{ij}p_{i}p_{j},
\end{equation}%
where $S$ and $\tau $  are  the Jacobi action and the affine
parameter respectively, along the geodesics. In the spherically symmetric spacetime,  the  motion  of photon  can be controlled by the following
Hamiltonian
\begin{equation}
H=\frac{1}{2}g^{ij}p_{i}p_{j}=0.  \label{EqHamiltonian}
\end{equation}
For simplicity reasons,  we consider the  photon motion on   the equatorial plane  $\theta =\frac{\pi }{2}$. Indeed,  one can get the Hamiltonian  equation of  the photon given by
\begin{equation}
(rf(r)p_{r})^2-r^2p_{t}^{2}+f(r)p_{t}^{2}=0
\label{EqNHa}
\end{equation}
where $E=-p_t$ and $L=p_\phi$ are the conserved total energy  and the conserved angular momentum of the photon, respectively.  It is recalled that it  has  been  used  the $p_\mu$  notation being  the  four-momentum.
Using the Hamiltonian-Jacobi  formalism, the
equations of motion can be formulated as
\begin{eqnarray}
\frac{dt }{d\tau}&=&\frac{E}{f(r)},  \notag \\
&&  \notag \\
\frac{dr }{d\tau} &=&\pm\sqrt{f(r)\left(\frac{E^{2}}{f(r)}- \frac{%
L^{2}}{r^{2}}\right) },  \notag \\
&&  \notag \\
\frac{d\phi }{d\tau} &=&-\frac{L}{r^{2}}.
\label{Eqmotion}
\end{eqnarray}%
Indeed, the   shape of a black hole is totally defined by the limit of its shadow  being the visible shape of the unstable circular orbits of  the photons.  To reach that,   one exploits the radial equation of motion. The latter  takes the  following form
\begin{equation}
\label{22}
\Big(\frac{dr}{d\tau}\Big)^2+V_{eff}(r)=0,
\end{equation}
where $V_{eff}(r)$ indicates the effective potential for a  radial particle motion. In particular, it is    given by
\begin{equation}
\label{23}
V_{eff}=f(r)\left( \frac{L^{2}}{r^{2}}-\frac{E^{2}}{f(r)}\right).
\end{equation}
The maximal  value of the effective potential   which  provides   the  circular orbits and  the unstable photons is  required by the following constraint
\begin{equation}
\label{24}
V_{eff}=\frac{d V_{eff}}{d r}\Big|_{r=r_p}=0.
\end{equation}
Using  Eq.\eqref{23}  and  Eq.\eqref{24}, one can obtain
\begin{equation}
\label{25}
V_{eff}|_{r=r_p}=\frac{d V_{eff}}{d r}\Big|_{r=r_p}= \left\{
    \begin{array}{ll}
        &f(r)\left( \frac{L^{2}}{r_p^{2}}-\frac{E^{2}}{f(r_p)}\right)=0, \\
        \\
        &L^2\left(\frac{r_pf'(r_p)-2f(r_p)}{r_p^3}\right)=0,
    \end{array}
\right.
\end{equation}
where one has used the notation $f'(r)=\frac{\partial f(r)}{\partial r}$. The photon sphere radius $r_p$ of  the quintessential Hayward-AdS black holes  corresponds  to the real and  the positive solution of the following constraint
\begin{equation}
\label{26}
r_pf'(r_p)-2f(r_p)=0.
\end{equation}
It has been remarked that the AdS backgrounds do not affect the  photon sphere radius $r_p$.
It follows from  Eq.(\ref{26})    that it is complicated to determine  $r_{p}$
analytically. However, we can   perform a  numerical  computation   to solve  the corresponding  equation. The orbit equation for the photon is obtained by considering the equation
\begin{equation}
\frac{dr}{d\phi}=\pm \frac{r^{2}}{L}\sqrt{f(r)\left(\frac{E^{2}}{f(r)}- \frac{%
L^{2}}{r^{2}}\right) }.
\label{Eqorbit}
\end{equation}
It is noted that   the photon orbit is constrained  by
\begin{equation}
 \frac{dr}{d\phi}\Big\vert_{r=R}=0.
 \end{equation}
Then,  the previous equation becomes
\begin{equation}
\frac{dr}{d\phi}=\pm r\sqrt{f(r)\left[\frac{r^{2}f(R)}{R^{2}f(r)} -1\right] }%
.  \label{EqTp}
\end{equation}
To get the desired equation, we should  consider a light ray sending from a static observer situated  at $r_{ob}$ and transmitting into the past with an angle $\alpha_{ob}$ with respect to the radial direction.
In this way,  one has
\begin{equation}
\cot\alpha_{ob}=\frac{\sqrt{g_{rr}}}{\sqrt{g_{\phi\phi}}}\frac{dr}{d\phi}{\Big{|}}_{r=r_{ob}}=\frac{1}{r\sqrt{f(r)}}\frac{dr}{d\phi}{\Big{|}}_{r=r_{ob}}.
\label{ff1}
\end{equation}
Exploiting Eq.\eqref{ff1},   one  gets
\begin{equation}
 \sin^{2}\alpha_{ob}=\frac{f(r_{ob})R^{2}}{r_{ob}^{2}f(R)}.
\label{effph}
\end{equation}
In this context,  one could  obtain  the  angular radius of the black hole shadow  by sending   $R$ to $r_{p}$  which is the circular orbit radius of the photon appearing in   Eq.\eqref{26}. Precisely,  the  shadow radius of the black hole observed by a static observer placed  at $r_{ob}$ has been found to be
\begin{equation}\label{shara}
r_{s}=r_{ob}\sin\alpha_{ob}=\left. R\sqrt{\frac{f(r_{ob})}{f(R)}}\right|_{R=r_{p}}.
\end{equation}
As the previous sections,  taking small values of the DE field  intensity, the radius  can be factorized as
\begin{equation}
\label{ }
r_{s}\sim r_{s}(c=0)+r_{s}(c)+O\left(c^2\right),
\end{equation}
where the involved terms are given by
\begin{eqnarray}
 r_{s}(c=0) & = & R\sqrt{\frac{g(r_{ob})}{g(R)}} \\
 r_{s}(c) & = & \frac{R\left(g(R)g'(r_{ob})-g(r_{ob})g'(R)\right)c}{2g(R)\sqrt{g(R)g(r_{ob})}}
\end{eqnarray}
with  $g(r)=f(r, c=0)$ and $g'=\frac{\partial f}{\partial c}$.
Following  \cite{Eiroa:2017uuq}, the  apparent shape of  the  shadow is obtained by using the celestial
coordinates $x$ and $y$. They are defined by
\begin{eqnarray}
x &=&\lim_{r_{0}\longrightarrow \infty }\left( -r_{0}^{2}\sin \theta _{0}%
\frac{d\phi }{dr}\Big\vert_{(r_{0},\theta _{0})}\right) ,  \notag \\
&&  \notag \\
y &=&\lim_{r_{0}\longrightarrow \infty }\left( r_{0}^{2}\frac{d\theta }{dr}%
\Big\vert_{(r_{0},\theta _{0})}\right).
\end{eqnarray}%
Fixing the value of  the state parameter to $\omega=-\frac{1}{3}$, the shadow geometries are illustrated in Fig.(\ref{f222}).
For  $Q=0.3$  and $\Lambda=-0.03$,  it has been remarked that    $c$ controls the size of the shadows.  It increases with   the  DE field intensity.
It has been observed that  the shadows of the  ordinary Hayward-AdS  black holes involve small radius compared  to the quintessential ones\cite{dark,dark1}. Taking   $Q=0.3$  and $c=0.3$,  we  remark similar behaviors in terms of the cosmological  constant.   An  examination shows that the flat solution  involves a large radius compared to the AdS  backgrounds.
For  $c=0.1$  and $\Lambda=-0.02$, however,  the charge  provides  a negligible effect on the  circular shadow  geometries. To go beyond such an analysis, we illustrate the radius variation in terms of  the quintessence field  intensity by  taking different values of  the  involved parameters. This is depicted  in Fig.(\ref{F223}).
For  $Q=0.3$  and $\Lambda=-0.09$,  it has been  observed that   the radius of the shadows for $\omega_q=-\frac{2}{3}$  are  larger than the ones associated with $\omega=-\frac{1}{3}$.   This implies  that  the state parameter could be considered as a parameter controlling the size of the shadow geometries.  By  increasing   the state parameter,  the shadow size decreases. Taking    $Q=0.3$  and  $\omega_q=-\frac{1}{3}$,   the cosmological  constant  increases the shadow radius.
In the range  $ 0< c<0.15$ with  $\omega_q=-\frac{1}{3}$ and $\Lambda=-0.02$,   the shadow size for  $Q=0.09$ is almost smaller  with respect to other  charge values. In the remaining range, the charge   has  a negligible effect on the  circular shadow   behaviors.
\subsection{Energy emission rate}
\begin{figure}[!ht]
		\centering
			\begin{tabbing}
			\hspace{5.4cm}\= \hspace{5.7cm}\=\kill
			\includegraphics[scale=.35]{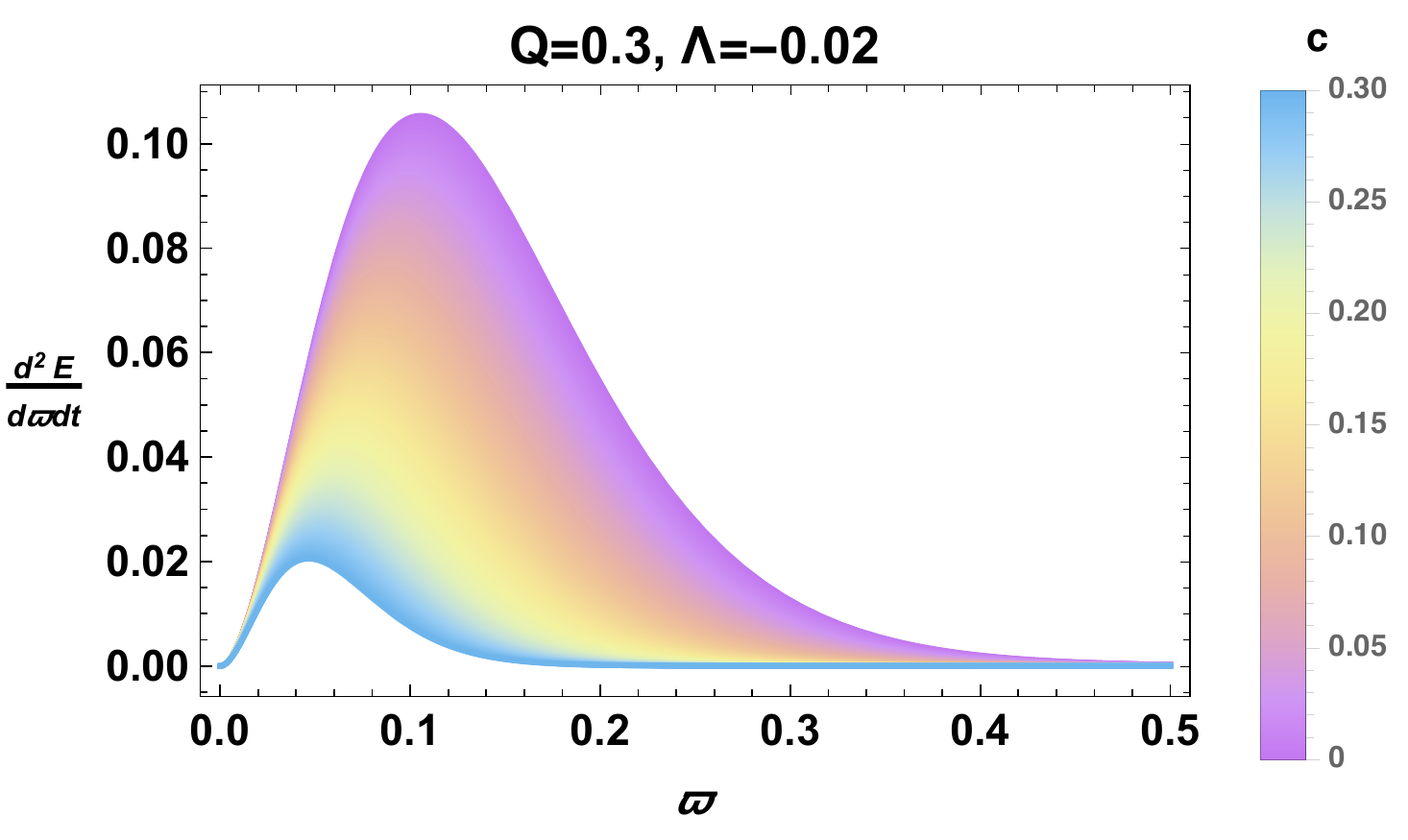} \>
			\includegraphics[scale=.35]{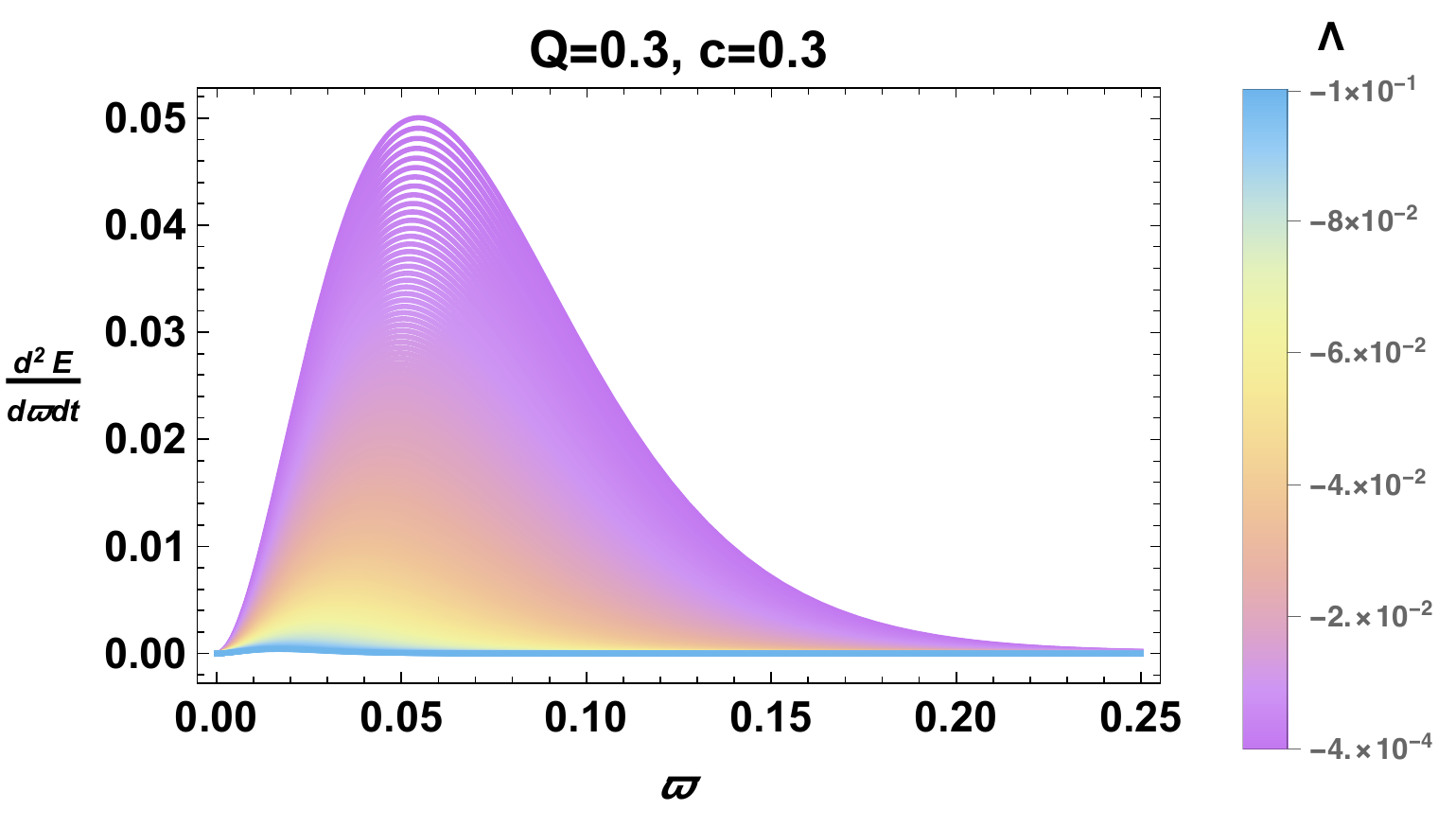} \>
			\includegraphics[scale=.35]{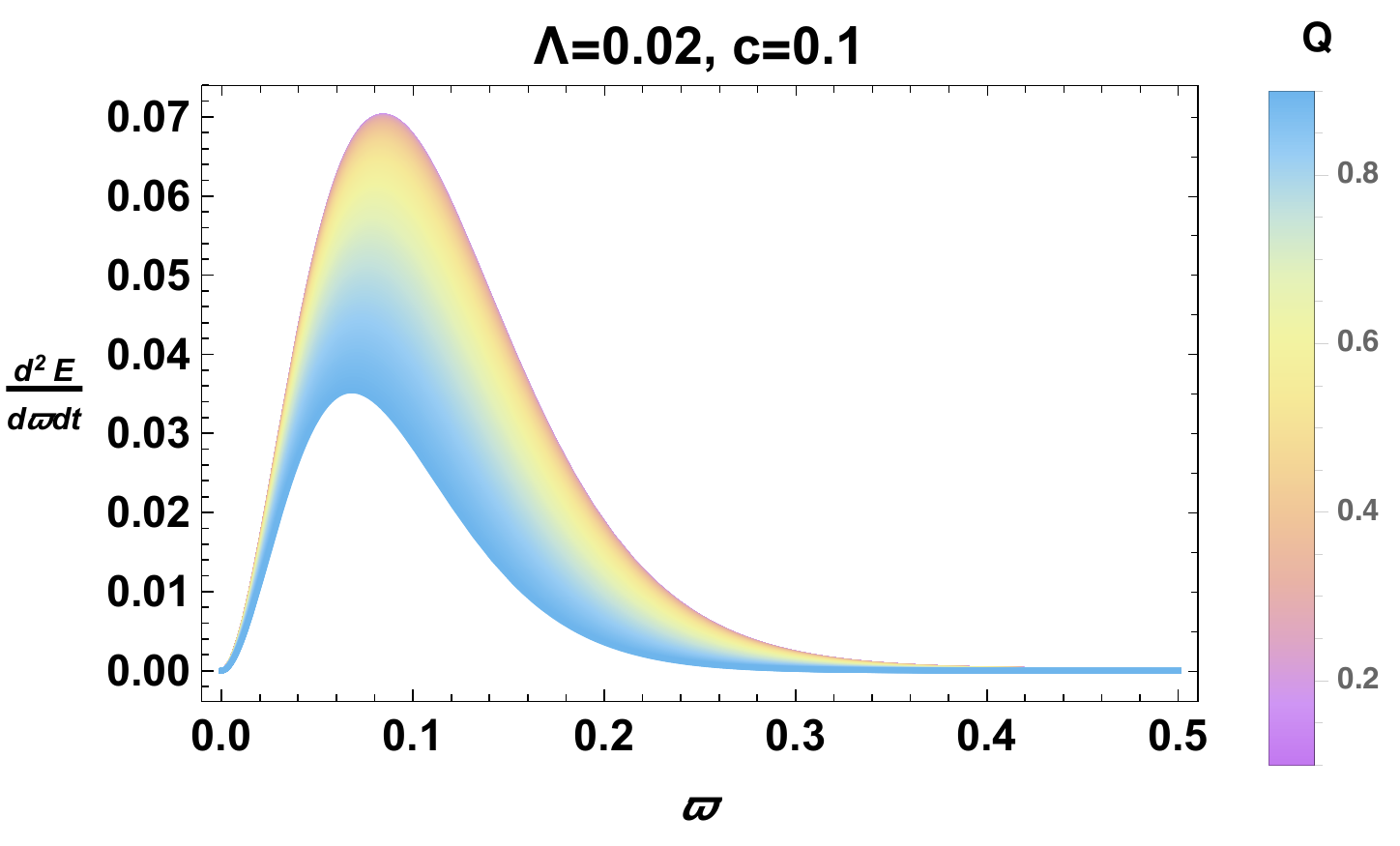} \\
		          \end{tabbing}
		          \vspace{-1cm}
\caption{{\it \footnotesize Energy emission rate for different  values of DE the intensity $c$, cosmological constant $\Lambda$ and charge $Q$, by takin  $\omega_q=-1/3$.}}
\label{FA}
\end{figure}

Here, we investigate the associated energy emission rate.  It is recalled  that near  the black hole horizons the   quantum fluctuations can  create and annihilate certain  pairs particles. In this  regard,  the positive energy particles  can escape through tunneling from the black hole, inside region where the Hawking radiation occurs.  This  phenomenon  is known as the  Hawking radiation  which causes   the black hole to evaporate in a certain period of time.  In what follows, we discuss   the  corresponding energy emission rate.  For a far distant observer,  the high energy absorption cross section  could approach to  the shadow of the  black hole.   At very high energy,   it has been  noted  that  the  absorption cross section of the black hole can  oscillate  to a limiting constant value $\sigma_{lim}=\pi r_s^2$.   Roughly, the  energy emission rate can be written as 
\begin{equation}
\label{72}
\frac{d^2 E(\varpi)}{d\varpi dt}=\frac{2\pi^{3} r_s^{2}\varpi^3}{e^{\frac{\varpi}{T_{{H}}}}-1},
\end{equation}
where $\varpi$    represents  the emission frequency \cite{Wei:2013kza}, and where   $T_{H}$ is the  associated  Hawking temperature.  For the studied model, one takes the  temperature given in Eq.(\ref{temp}).
The energy emission rate is  represented  in Fig.(\ref{FA}) as a function of $\varpi$ by varying $c$, $\Lambda$ and $Q$ parameters.
It is  follows from  Fig.(\ref{FA}) that, when DE is present,  the energy emission rate is lower. This  indicates that the black hole evaporation process is slow. Moreover, we  get an even slower radiation process by  increasing (decreasing)   the intensity $c$ and the charge $Q$ (the cosmological constant $\Lambda$ parameter and the parameter $\omega_q$).  It has been  remarked that   the emission rate of   the Hayward-AdS black holes is slow compared  to  other  back hole solutions \cite{aaa, bbb, ccc}.
\section{Conclusions and discussions}
In this paper, we have investigated the thermodynamic and  the optical behaviors  of the  quintessential  Hayward-AdS black holes in four dimensions.  For the  thermodynamic aspect,  we have  first  reconsidered the study of  the critical behaviors   of the  ordinary black hole solutions. Concretely, we  have found that the equation of state  provides  a universal  ratio  given by $\chi_0=\frac{P_cv_c}{T_c}=\frac{27-3\sqrt{6}}{50}$.   For certain values of the DE state parameter $\omega_q$, this new number has been modified by a contribution $\xi_{\omega_q}(c,Q)$  depending on the  DE field intensity and the charge. Then, we   have  studied  the DE effect on  the heat engine behaviors of such  Hayward AdS  black hole solutions by taking certain  known values of $\omega_q=-1,-\frac{2}{3},-\frac{1}{3} $.  For  the optical aspect,  we have inspected   the influence of DE  by considering  shadow  geometries using one dimensional real space. Concretely,  we  have obtained  that  the associated radius  has been  changed   by DE contributions.   For  different values of $\omega_q$ the efficiency  and the ratio $\eta/\eta_{car}$ go to $1$ by increasing the normalization factor of the quintessence $c$. This  can be confirmed   by  the  fact that the quintessence field cools the black hole making it more efficient.  Then, we have shown that  the Hayward-AdS black holes involve small radius compared to the quintessential ones, in connection with EHT, such new behaviors could take  good places. However, the variation of the black hole shadow radius as a  function of  the parameter $Q$  related to the total magnetic charge $q_m$ is neglected. This could  reveal  that  the parameter $Q$ does not  affect the black hole shadows.
\\
This work comes with some open questions.   A natural  question may concern the implementation of other  quantities  including the  rotation parameter.   It could be possible to get certain  conditions for dark energy models  which could be considered as  an alternative cosmological model to $\Lambda$CMD. Higher dimensional models could be an interesting  investigation in future works.

\section*{Acknowledgements}
The authors would like to thank H. Belmahi, A. El Balali, W. El Hadri, Y. Hassouni  and E.
Torrente Lujano for discussions on related topics. We are also grateful to the anonymous referee for careful reading of our manuscript, insightful comments, and suggestions, which improve the quality of the present paper significantly.
This
work is partially supported by the ICTP through AF.


\begin{thebibliography}{99}

\bibitem{Kastor:2009wy}
D.~Kastor, S.~Ray and J.~Traschen, {\it Enthalpy and the Mechanics of AdS Black Holes}, Class. Quant. Grav. \textbf{26} (2009) 195011, arXiv:0904.2765.

\bibitem{Dolan:2011xt}
B.~P.~Dolan, {\it Pressure and volume in the first law of black hole thermodynamics}, Class. Quant. Grav. \textbf{28} (2011) 235017, arXiv:1106.6260.

\bibitem{Kubiznak:2012wp}
D.~Kubiznak and R.~B.~Mann, {\it P-V criticality of charged AdS black holes}, JHEP \textbf{07} (2012) 033, arXiv:1205.0559.

\bibitem{Altamirano:2013ane}
N.~Altamirano, D.~Kubiznak and R.~B.~Mann, {\it Reentrant phase transitions in rotating anti\textendash{}de Sitter black holes},
Phys. Rev. D \textbf{88} 10 (2013) 101502, arXiv:1306.5756.
 
\bibitem{ma1} 
A. Awad,  G. G.L. Nashed, {\it Generalized teleparallel cosmology and initial singularity crossing},J.C.A.P \textbf{02}(2017) 046, {\tt arXiv:1701.06899}.

\bibitem{ma2} 
T.  Shirafuji,  G. G. L.  Nashed, {\it Energy and Momentum in the Tetrad Theory of Gravitation }, P. T. Phys, \textbf{6} (1997) 98, {\tt arXiv:gr-qc/9711010}.

\bibitem{ma3} 
E. Elizalde, G.G.L. Nashed, S. Nojiri, S.D. Odintsov, {\it Spherically symmetric black holes with electric and magnetic charge in extended gravity: Physical properties, causal structure, and stability analysis in Einstein’s and Jordan’s frames}, Eur. Phys. J. C \textbf{2}, (2020) 80, {\tt arXiv:2001.11357}. 

\bibitem{ma4} 
A. Awad, W. El Hanafy, G.G.L. Nashed, S.D.Odintsov, V.K. Oikonomou, {\it Constant-roll Inflation in $f(T)$
Teleparallel Gravity}, J. C. A. P, \textbf{07} (2018)26, {\tt arXiv: 1710.00682}.

\bibitem{ma5}
 G. G. L. Nashed, {\it Brane World black holes in Teleparallel Theory Equivalent to General Relativity and their Killing vectors, Energy, Momentum and Angular-Momentum}, Chinese Phys. B \textbf{19} (2010) 020401, {\tt arXiv:0910.5124}.

\bibitem{ma6}
 W. El Hanafy and G. G. L. Nashed, {\it Exact Teleparallel Gravity of Binary Black Holes}, Astrophys Space Sci \textbf{361},(2016) 68, {\tt arXiv:1507.07377}.

\bibitem{ma7} 
G. G.L. Nashed, {\it Schwarzschild solution in extended teleparallel gravity},E.P.L, \textbf{105}, 10001, {\tt arXiv:1501.00974}


\bibitem{Kubiznak:2015bya}
D.~Kubiznak and F.~Simovic, {\it Thermodynamics of horizons: de Sitter black holes and reentrant phase transitions}, Class. Quant. Grav. \textbf{33} 24 (2016) 245001, arXiv:1507.08630.

\bibitem{Altamirano:2013uqa}
N.~Altamirano, D.~Kubiz\v{n}\'ak, R.~B.~Mann and Z.~Sherkatghanad, {\it Kerr-AdS analogue of triple point and solid/liquid/gas phase transition}, Class. Quant. Grav. \textbf{31} (2014) 042001, arXiv:1308.2672.

\bibitem{Hennigar:2016xwd}
R.~A.~Hennigar, R.~B.~Mann and E.~Tjoa, {\it Superfluid Black Holes}, Phys. Rev. Lett. \textbf{118} 2 (2017) 021301, arXiv:1609.02564.

\bibitem{Bronnikov:2000vy}
K.~A.~Bronnikov, {\it Regular magnetic black holes and monopoles from nonlinear electrodynamics}, Phys. Rev. D \textbf{63} (2001) 044005, arXiv:gr-qc/0006014.

\bibitem{Dymnikova:2004zc}
I.~Dymnikova,
{\it Regular electrically charged structures in nonlinear electrodynamics coupled to general relativity}, Class. Quant. Grav. \textbf{21} (2004) 4417, arXiv:gr-qc/0407072.

\bibitem{Polchinski:1989ae}
J.~Polchinski, {\it Decoupling Versus Excluded Volume or Return of the Giant Wormholes}, Nucl. Phys. B \textbf{325} (1989) 619.


\bibitem{Hayward:2005gi}
S.~A.~Hayward, {\it Formation and Evaporation of Regular Black Holes}, Phys. Rev. Lett. \textbf{96} (2006) 031103,  arXiv:gr-qc/0506126.

\bibitem{Frolov:1989pf}
V.~P.~Frolov, M.~A.~Markov and V.~F.~Mukhanov, {\it Through a Black Hole Into a New Univers?}, Phys. Lett. B \textbf{216} (1989) 272.

\bibitem{Mukhanov:1991zn}
V.~F.~Mukhanov and R.~H.~Brandenberger, {A Nonsingular Universe}, Phys. Rev. Lett. \textbf{68},  (1992)1969.


\bibitem{Frolov:2016pav}
V.~P.~Frolov, {\it Notes on nonsingular models of black holes}, Phys. Rev. D \textbf{94} 10 (2016) 104056, arXiv:1609.01758.

\bibitem{Modesto:2006mx}
L.~Modesto, {\it Black hole interior from loop quantum gravity}, Adv. High Energy Phys. \textbf{2008} (2008) 459290, arXiv:gr-qc/0611043.

\bibitem{Bertone:2016nfn}
G.~Bertone and D.~Hooper,
{\it History of dark matter}, Rev. Mod. Phys. \textbf{90}, 4 (2018) 045002, arXiv:1605.04909.

\bibitem{Huterer:2017buf}
D.~Huterer and D.~L.~Shafer,
{\it Dark energy two decades after: Observables, probes, consistency tests}, Rept. Prog. Phys. \textbf{81}1 (2018) 016901,arXiv:1709.01091.

\bibitem{Gannouji:2019mph}
R.~Gannouji, {\it A Primer on Modified Gravity}, Int. J. Mod. Phys. D \textbf{28} 05 (2019) 1942004.
\bibitem{180}
A. Belhaj, A. El Balali, W. El Hadri, H. El Moumni, M. B. Sedra, Dark energy effects
on charged and rotating black holes, Eur. Phys. J. Plus 134(9) (2019) 422.

\bibitem{Caldwell:1997ii}
R.~R.~Caldwell, R.~Dave and P.~J.~Steinhardt, {\it Cosmological imprint of an energy component with general equation of state}, Phys. Rev. Lett. \textbf{80}, (1998) 1582, arXiv:astro-ph/9708069.


\bibitem{Uniyal:2014paa}
R.~Uniyal, N.~Chandrachani Devi, H.~Nandan and K.~D.~Purohit,
{\it Geodesic Motion in Schwarzschild Spacetime Surrounded by Quintessence}, Gen. Rel. Grav. \textbf{47} 2 (2015) 16, arXiv:1406.3931.

\bibitem{Konoplya:2019sns}
R.~A.~Konoplya,
{\it Shadow of a black hole surrounded by dark matter}, Phys. Lett. B \textbf{795}, (2019) 6, arXiv:1905.00064.


\bibitem{RF1}
N.~A.~Bahcall, J.~P.~Ostriker, S.~Perlmutter and P.~J.~Steinhardt,
{\it The Cosmic triangle: Assessing the state of the universe},
Science \textbf{284} (1999) 1481, {\tt arXiv:astro-ph/9906463}.

\bibitem{RF2}
V.~Sahni and A.~A.~Starobinsky, {\it The Case for a positive cosmological Lambda term}, Int. J. Mod. Phys. D \textbf{9} (2000) 373, {\tt arXiv:astro-ph/9904398}.

\bibitem{Kiselev:2002dx}
V.~V.~Kiselev, {Quintessence and black holes}, Class. Quant. Grav. \textbf{20},  (2003) 1187, arXiv:gr-qc/0210040.

\bibitem{Chen:2008ra}
S.~Chen, B.~Wang and R.~Su, {\it Hawking radiation in a $d$-dimensional static spherically-symmetric black Hole surrounded by quintessence}, Phys. Rev. D \textbf{77}, (2008) 124011, arXiv:0801.2053.



\bibitem{Chen:2012mva}
S.~Chen, Q.~Pan and J.~Jing,{\it Holographic superconductors in quintessence AdS black hole spacetime}, Class. Quant. Grav. \textbf{30} (2013) 145001, arXiv:1206.2069.

\bibitem{Chabab:2017xdw}
M.~Chabab, H.~El Moumni, S.~Iraoui, K.~Masmar and S.~Zhizeh, {More Insight into Microscopic Properties of RN-AdS Black Hole Surrounded by Quintessence via an Alternative Extended Phase Space},Int. J. Geom. Meth. Mod. Phys. \textbf{15}, 10 (2018) 1850171, arXiv:1704.07720.

\bibitem{Belhaj:2020rdb}
A.~Belhaj, M.~Benali, A.~El Balali, H.~El Moumni and S.~E.~Ennadifi, {\it Deflection angle and shadow behaviors of quintessential black holes in arbitrary dimensions}, Class. Quant. Grav. \textbf{37} 21 (2020)
 215004, arXiv:2006.01078.


\bibitem{JT1}
\"O.~\"Okc\"u and E.~Ayd\i{}ner,
{\it Joule\textendash{}Thomson expansion of the charged AdS black holes},
Eur. Phys. J. C \textbf{77} (2017) no.1, 24
[arXiv:1611.06327 [gr-qc]].




\bibitem{JT2}
M.~Chabab, H.~El Moumni, S.~Iraoui, K.~Masmar and S.~Zhizeh,
{\it Joule-Thomson Expansion of RN-AdS Black Holes in $f(R)$ gravity},
LHEP \textbf{02} (2018), 05
[arXiv:1804.10042 [gr-qc]].


\bibitem{Belhaj:2015hha}
A.~Belhaj, M.~Chabab, H.~El Moumni, K.~Masmar, M.~B.~Sedra and A.~Segui,
{\it On Heat Properties of AdS Black Holes in Higher Dimensions},
JHEP \textbf{05} (2015), 149
[arXiv:1503.07308 [hep-th]].


\bibitem{27}
C.~V.~Johnson, {\it Holographic Heat Engines}, Class. Quant. Grav. \textbf{31} (2014) 205002, arXiv:1404.5982.


\bibitem{28}
M.~Zhang, C.~M.~Zhang, D.~C.~Zou and R.~H.~Yue, {\it $P-V$ criticality and Joule-Thomson Expansion of Hayward-AdS black holes in 4D Einstein-Gauss-Bonnet gravity}, arXiv:2102.04308.



\bibitem{NPBx}
H.~El Moumni and K.~Masmar,
{\it Regular AdS black holes holographic heat engines in a benchmarking scheme},
Nucl. Phys. B \textbf{973} (2021), 115590.



\bibitem{29}
Y-L.~Huang and S.~Guo, {Thermodynamic of the charged accelerating AdS black hole: P-V critical and Joule-Thomson expansion}, arXiv:2009.09401 .



\bibitem{30}
H.~Liu and X.~H.~Meng, {\it Effects of dark energy on the efficiency of charged AdS black holes as heat engines}, Eur. Phys. J. C \textbf{77} 8 (2017) 556, arXiv:1704.04363.


\bibitem{shad1}
X.~X.~Zeng and H.~Q.~Zhang,
{\it Influence of quintessence dark energy on the shadow of black hole},
Eur. Phys. J. C \textbf{80} (2020) no.11, 1058,
arXiv:2007.06333 [gr-qc].


\bibitem{shad2}
S.~U.~Khan and J.~Ren,
{\it Shadow cast by a rotating charged black hole in quintessential dark energy},
Phys. Dark Univ. \textbf{30} (2020) 100644,arXiv:2006.11289 [gr-qc].

\bibitem{shad3}
O.~Pedraza, L.~A.~L\'opez, R.~Arceo and I.~Cabrera-Munguia,
{\it Geodesics of Hayward black hole surrounded by quintessence},
Gen. Rel. Grav. \textbf{53} (2021) 24, arXiv:2008.00061 [gr-qc].



\bibitem{shad4}
A.~Abdujabbarov, B.~Toshmatov, Z.~Stuchl\'\i{}k and B.~Ahmedov,
{\it Shadow of the rotating black hole with quintessential energy in the presence of plasma},
Int. J. Mod. Phys. D \textbf{26} (2016)  1750051, arXiv:1512.05206 [gr-qc].


\bibitem{shad5}
A.~Abdujabbarov, M.~Amir, B.~Ahmedov and S.~G.~Ghosh,
{\it Shadow of rotating regular black holes},
Phys. Rev. D \textbf{93} (2016) 104004, 
arXiv:1604.03809 [gr-qc].


\bibitem{31}
Z.~Y.~Fan and X.~Wang,
{\it Construction of Regular Black Holes in General Relativity}, Phys. Rev. D \textbf{94} 12 (2016) 124027, arXiv:1610.02636.

\bibitem{Fan}
Z.~Y.~Fan,
{\it Critical phenomena of regular black holes in anti-de Sitter space-time},
Eur. Phys. J. C \textbf{77} (2017) no.4, 266, 
arXiv:1609.04489 [hep-th].

\bibitem{thermo1}
D.~Kubiznak and R.~B.~Mann,
{\it P-V criticality of charged AdS black holes},
JHEP \textbf{07} (2012) 033,
arXiv:1205.0559 [hep-th].



\bibitem{thermo2}
A.~Belhaj, M.~Chabab, H.~El Moumni and M.~B.~Sedra,
{\it On Thermodynamics of AdS Black Holes in Arbitrary Dimensions},
Chin. Phys. Lett. \textbf{29} (2012) 100401, 
arXiv:1210.4617 [hep-th].



\bibitem{Carter:1968rr}
B.~Carter,
{\it Global structure of the Kerr family of gravitational fields},
Phys. Rev. \textbf{174} (1968), 1559-1571



\bibitem{Chandrasekhar}
S. Chandrasekhar, {\it The Mathematical Theory of Black Holes}, (Oxford University Press,
1998).

\bibitem{Eiroa:2017uuq}
E.~F.~Eiroa and C.~M.~Sendra,
{\it Shadow cast by rotating braneworld black holes with a cosmological constant},
Eur. Phys. J. C \textbf{78} (2018)  91,
arXiv:1711.08380 [gr-qc].
\bibitem{dark}
A. Belhaj, A. El Balali, W. El Hadri, M. A. Essebani, M. B. Sedra and A. Segui, {\it KerrAdS Black Hole Behaviors from Dark Energy}, Int. Jour. of Mod. Phys. D{\bf 29} (09) (2020)
2050069.
\bibitem{dark1}
A. Belhaj, A. El Balali, W. El Hadri, Y. Hassouni, E. Torrente-Lujan, {\it
Phase transition and shadow behaviors of quintessential black holes in M-theory/superstring inspired models},  Int.J.Mod.Phys. A {\bf 36} (2021) 2150057.
\bibitem{Wei:2013kza}
S.~W.~Wei and Y.~X.~Liu,
{\it Observing the shadow of Einstein-Maxwell-Dilaton-Axion black hole},
JCAP \textbf{11} (2013) 063,
arXiv:1311.4251 [gr-qc].

\bibitem{ddd} 
Z-Y. Fan and X. Wang, {\it Construction of Regular Black Holes in General Relativity}, Phys. Rev. D 94, 124027 (2016), {\tt arXiv:1610.02636 [gr-qc]}.

\bibitem{aaa}
A. Belhaj, M. Benali, A. El Balali, W. El Hadri, H. El Moumni, {\it Shadows of Charged and Rotating Black Holes with a Cosmological Constant}, Gen. Rel. and Qua. Cos. (2020), {\tt arXiv:2007.09058"[gr-qc]}.
  \bibitem{bbb}  
    S. W. Wei, Y. X. Liu, {\it Observing the shadow of Einstein-Maxwell-Dilaton-Axion black hole},
JCAP 11 (2013)063, arXiv:1311.4251 [gr-qc].
  \bibitem{ccc}
  S. V. M. C. B. Xavier, Pedro V. P. Cunha, Luıs C. B. Crispino, Carlos A. R. Herdeiro, \textit{Shadows
of charged rotating black holes: Kerr-Newman versus Kerr-Sen}, arXiv:2003.14349.


\end{thebibliography}
\end{document}